\documentclass[amsmath,floatfix,twocolumn,superscriptaddress]{revtex4-1}
\usepackage{subfigure}
\usepackage{siunitx}
\usepackage{amssymb}
\usepackage{amsmath}
\usepackage{graphicx}
\usepackage{array}
\usepackage{dcolumn}
\usepackage{psfrag}
\usepackage{bm}
\usepackage{color}
\usepackage{multirow}

\begin{document}

\title{Machine Learning-Guided Discovery of Temperature-Induced Solid-Solid Phase Transitions in Inorganic Materials}

\author{Cibrán López}
    \affiliation{Group of Characterization of Materials, Departament de F\'{i}sica, Universitat Polit\`{e}cnica de 
    Catalunya, Campus Diagonal Bes\`{o}s, Av. Eduard Maristany 10--14, 08019 Barcelona, Spain}
    \affiliation{Research Center in Multiscale Science and Engineering, Universitat Politècnica de Catalunya, Campus 
    Diagonal-Bes\`{o}s, Av. Eduard Maristany 10--14, 08019 Barcelona, Spain}

\author{Joshua Ojih}
    \affiliation{Department of Mechanical Engineering, University of South Carolina, Columbia, SC 29208, USA}

\author{Ming Hu}
    \affiliation{Department of Mechanical Engineering, University of South Carolina, Columbia, SC 29208, USA}

\author{Josep-Llu\'is Tamarit}
    \affiliation{Group of Characterization of Materials, Departament de F\'{i}sica, Universitat Polit\`{e}cnica de 
    Catalunya, Campus Diagonal Bes\`{o}s, Av. Eduard Maristany 10--14, 08019 Barcelona, Spain}
    \affiliation{Research Center in Multiscale Science and Engineering, Universitat Politècnica de Catalunya, Campus 
    Diagonal-Bes\`{o}s, Av. Eduard Maristany 10--14, 08019 Barcelona, Spain}

\author{Edgardo Saucedo}
	\affiliation{Research Center in Multiscale Science and Engineering, Universitat Politècnica de Catalunya, 
	 Campus Diagonal-Bes\`{o}s, Av. Eduard Maristany 10--14, 08019 Barcelona, Spain}
	\affiliation{Micro and Nanotechnologies Group, Emerging Thin Film Photovoltaics Lab, Departament d’Enginyeria 
	Electr\`{o}nica, Universitat 
	Polit\`{e}cnica de Catalunya, Campus Diagonal Bes\`{o}s, Av. Eduard Maristany 10-14, 08019 Barcelona, Spain.}

\author{Claudio Cazorla}
    \affiliation{Group of Characterization of Materials, Departament de F\'{i}sica, Universitat Polit\`{e}cnica de 
    Catalunya, Campus Diagonal Bes\`{o}s, Av. Eduard Maristany 10--14, 08019 Barcelona, Spain}
    \affiliation{Research Center in Multiscale Science and Engineering, Universitat Politècnica de Catalunya, Campus 
    Diagonal-Bes\`{o}s, Av. Eduard Maristany 10--14, 08019 Barcelona, Spain}

\begin{abstract}
\textbf{Abstract.}~Predicting solid-solid phase transitions remains a long-standing challenge in materials science. 
Solid-solid transformations underpin a wide range of functional properties critical to energy conversion, information 
storage, and thermal management technologies. However, their prediction is computationally intensive due to the need 
to account for finite-temperature effects. Here, we present an uncertainty-aware machine-learning-guided framework 
for high-throughput prediction of temperature-induced polymorphic phase transitions in inorganic crystals. By combining 
density functional theory calculations with graph-based neural networks trained to estimate vibrational free energies, 
we screened a curated dataset of $\sim 50,000$ inorganic compounds and identified over $2,000$ potential solid-solid 
transitions within the technologically relevant temperature interval $300$--$600$~K. Among our key findings, we uncover 
numerous phase transitions exhibiting large entropy changes ($> 300$~J~K$^{-1}$~kg$^{-1}$), many of which occur near 
room temperature hence offering strong potential for solid-state cooling applications. We also identify $21$ compounds 
that exhibit substantial relative changes in lattice thermal conductivity ($20$--$70$\%) across a phase transition, 
highlighting them as promising thermal switching materials. Validation against experimental observations and first-principles 
calculations supports the robustness and predictive power of our approach. Overall, this work establishes a scalable 
route to discover functional phase-change materials under realistic thermal conditions, and lays the foundation for 
future high-throughput studies leveraging generative models and expanding open-access materials databases.
\\

{\bf Keywords:} solid-solid phase transitions, machine learning, first-principles calculations, high-throughput screening, 
	        solid-state cooling, thermal switching, inorganic crystals  
\end{abstract}

\maketitle

\section{Introduction}
\label{sec:intro}
Solid-solid phase transitions play a crucial role in governing a wide range of physical phenomena and underpin 
numerous technological applications. For example, polymorphic transitions in crystalline materials influence mechanical 
properties in metallurgy, as seen in the martensitic transformation of steel, which enhances its strength and toughness. 
In functional materials, phase transitions enable key applications, such as the ferroelectric-paraelectric transition 
in BaTiO$_{3}$, which is essential for capacitors, or the shape-memory effect in NiTi alloys used in medical devices. 
The ability to control and predict these phase transitions is fundamental for optimizing material performance in areas 
ranging from structural engineering to electronic and energy technologies.

During a phase transition a material undergoes a transformation between different states of matter (e.g., solid to liquid)
or structural phases (e.g., martensitic transformation) due to changes in external conditions such as temperature, pressure,
or composition \cite{pt1,pt2,pt3,pt4}. These phase transitions are broadly classified into first-order
and second-order types: first-order transitions involve latent heat and discontinuous changes in order parameters, as seen
in melting, while second-order transitions occur continuously without latent heat, as in ferromagnetic to
paramagnetic transitions. The nature of a phase transition is governed by underlying thermodynamic and atomistic mechanisms,
which can be described using concepts from quantum physics, statistical mechanics and symmetry breaking. Understanding phase 
transitions is essential for predicting crystal properties and designing advanced functional materials.

\begin{figure*}[t]
    \centering
    \includegraphics[width=0.8\linewidth]{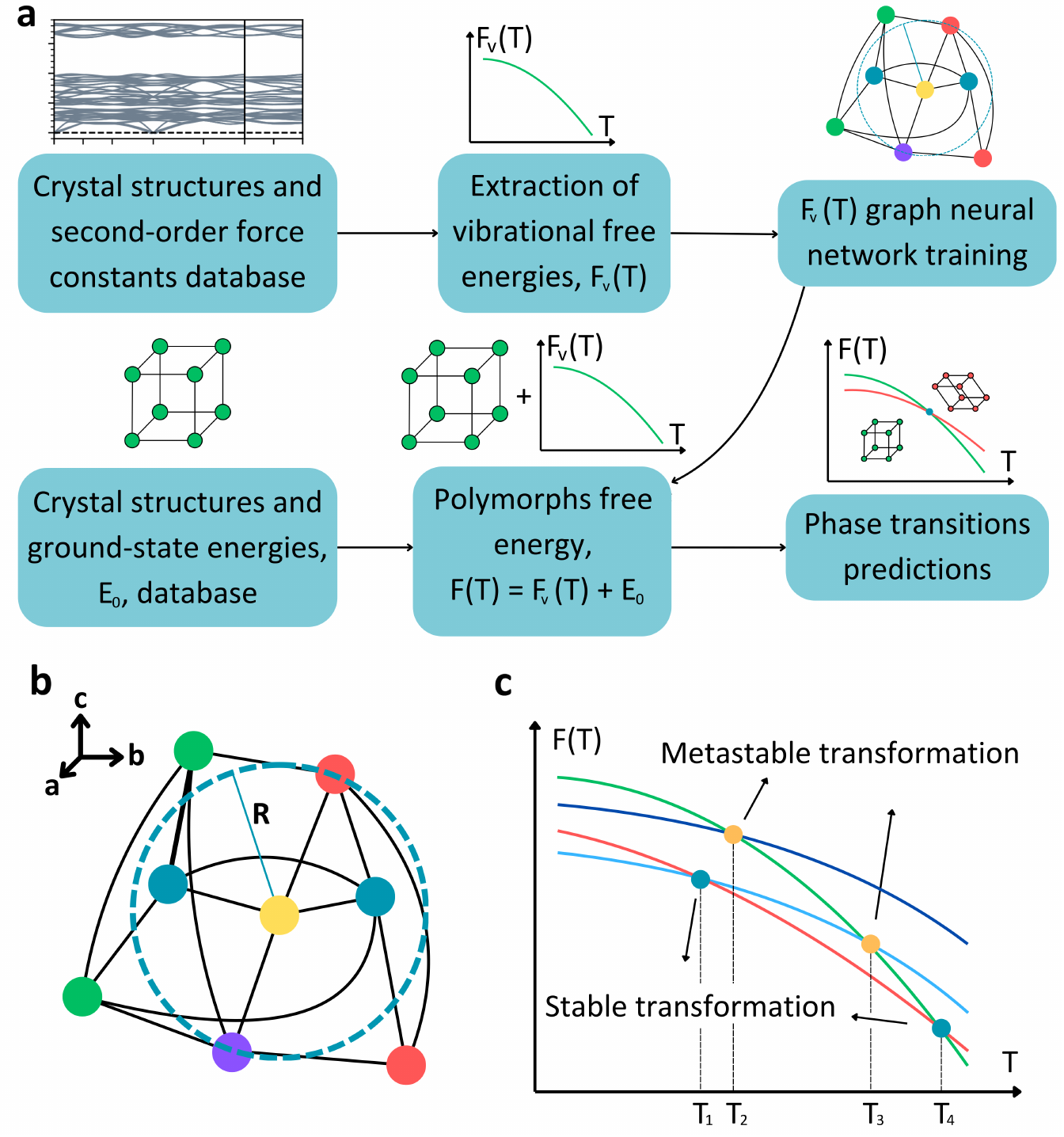}
	\caption{\textbf{Machine learning-guided high-throughput approach for prediction of solid-solid phase transitions 
	in inorganic crystals.}
	\textbf{a.}~Sketch of the employed computational workflow relying on DFT and vibrational free-energy calculations.
	\textbf{b.}~Diagram of a graph encoding a crystal structure. The graph is characterized by nodes (dots), edges 
	(lines), and a cutoff distance ($R$, dashed line).
	\textbf{c.}~Determination of temperature-induced phase transitions, both stable and metastable, based on 
		    free-energy curve calculations.} 
    \label{fig1}
\end{figure*}

Phase transitions can occur under a broad range of thermodynamic conditions, including variations in temperature, 
pressure, and external fields, making their precise characterization inherently complex. Crucially, the occurrence 
and nature of these transitions cannot always be predicted solely from a material's chemical composition or atomic 
structure, as they often arise from intricate many-body interactions and thermal effects. This lack of predictability 
poses a significant challenge for the rational design of materials as, for example, unexpected phase transitions may 
hinder or alter their intended functionality. Thus, the difficulty in foreseeing phase behavior limits the discovery 
of novel physical phenomena and constrains the development of advanced technological applications.

Despite the challenges in predicting phase transitions, several well-established theoretical and computational 
approaches have been developed to accurately model and anticipate solid-solid phase transformations \cite{met1,met2,met3,met4}. 
Methods such as free-energy calculations, Monte Carlo and molecular dynamics simulations enable the quantitative 
evaluation of phase stability by capturing the interplay between enthalpic and entropic contributions. However, 
achieving reliable predictions requires extremely high numerical accuracy, as small errors in energy estimations can 
lead to significant deviations in transition temperatures and stability trends. Consequently, these techniques demand 
rigorous computational protocols to ensure the precise determination of phase behavior in materials.

Furthermore, the computational cost associated with free-energy calculations and molecular dynamics simulations is 
substantial, as these methods require extensive sampling and highly accurate energy and forces evaluations. Consequently, 
their application is often limited to specific case studies rather than large-scale, high-throughput screenings of 
extensive material datasets. This technical constraint significantly hinders the systematic exploration of phase behavior 
across diverse material families, limiting the discovery of novel physical phenomena and emergent functionalities. 
As a result, promising materials with transformative technological potential may remain unexplored due to the prohibitive 
computational demands of current predictive approaches.

In this study, we have developed a computational framework that combines first-principles calculations with 
machine-learning techniques to overcome the limitations of conventional phase transition predictions. Specifically, 
we integrate density functional theory (DFT) energy calculations with uncertainty-aware machine-learning-assisted 
quasi-harmonic free-energy evaluations, enabling the systematic and high-throughput prediction of temperature-induced 
solid-solid phase transitions. By applying this approach to a dataset of approximately $50,000$ inorganic materials, 
we have identified around $2,000$ polymorphic phase transitions, some of which present potential technological relevance. 
Using these predictions, we propose novel materials for two distinct thermal management applications: solid-state cooling 
and thermal conductivity switching. Our findings not only identify promising phase transitions for energy-related 
technologies but also establish a scalable and efficient computational strategy for materials discovery under realistic 
finite-temperature conditions.

\section{Computational Approach}
\label{sec:approach}
We present a novel high-throughput methodology for predicting temperature-induced solid-solid phase transitions in 
crystals. By integrating first-principles DFT calculations with uncertainty-aware, machine-learning-assisted quasi-harmonic 
free-energy evaluations, this approach enables a robust and reliable prediction of vibrational free energies.

\subsection{Quasi-harmonic free-energy formalism} 
\label{subsec:free-energy}
In the quasi-harmonic (QH) approximation \cite{qh1,qh2,qh4}, the Helmholtz free energy of a nonmetallic and
nonmagnetic crystal, $F$, is expressed as a function of volume ($V$) and temperature ($T$) like:  
\begin{equation}  
        F(V,T) = E_{0}(V) +  F_{\rm v} (V,T)~,  
\label{eq:fener}  
\end{equation}  
where $E_{0}$ is the static lattice energy at zero temperature, and $F_{\rm v}$ is the vibrational free energy
(Fig.\ref{fig1}a). The static lattice energy can be directly computed using first-principles methods such as 
density functional theory (DFT), $E_{0} = E_{\rm DFT}$. Similarly, the vibrational free energy can be obtained 
from the crystal's phonon frequencies, $\omega_{\boldsymbol{q}s}(V)$, as:  
\begin{equation}  
        F_{\rm v} (V,T) = k_{B} T \sum_{\boldsymbol{q}s} \ln\left[ 2\sinh \left(\frac{\hbar  
        \omega_{\boldsymbol{q}s}}{2k_{\rm B}T} \right) \right]~,  
\label{eq:fharm}  
\end{equation}  
where the summation extends over the reciprocal-space vectors in the Brillouin zone, $\boldsymbol{q}$, and 
phonon branches, $s$. It is important to note that the QH approach is applicable only to vibrationally stable 
systems with strictly positive phonon frequencies, as imaginary (or negative) frequencies render $F_{\rm v}$ 
ill-defined.  

Once $F$ is known, any thermodynamic quantity can be derived from it. For example, the pressure is given by 
$P = - \left( \partial F/\partial V \right)_{T}$, and the entropy by $S = - \left( \partial F/\partial T \right)_{V}$. 
At zero pressure, a phase transition between two polymorphs, $A$ and $B$, occurs at the transition temperature 
$T_{t}$ when their Helmholtz free energies are equal:  
\begin{equation}  
    F_{A} (V_{A}, T_{t}) = F_{B} (V_{B}, T_{t})~,  
\label{eq:trans}
\end{equation}  
and the volume of the two polymorphs fulfill the condition $P_{A} (V_{A}) = P_{B} (V_{B}) = 0$. When $F_{A}$ 
and $F_{B}$ are known, these conditions enable the numerical determination of $T_{t}$ (Fig.\ref{fig1}c). 

In this study, we apply the free-energy criterion in Eq.(\ref{eq:trans}) to estimate phase transition temperatures, 
assuming that the equilibrium volume of the crystals remains constant with temperature (i.e., neglecting thermal 
expansion effects). A phase transition can occur between the most stable phase and the lowest free-energy metastable 
polymorph, in which case we refer to it as a ``stable'' phase transition. Conversely, when a transition occurs 
between two metastable polymorphs, we refer to it as a ``metastable'' phase transition (Fig.\ref{fig1}c).

\begin{figure*}[t]
    \centering
    \includegraphics[width=0.9\linewidth]{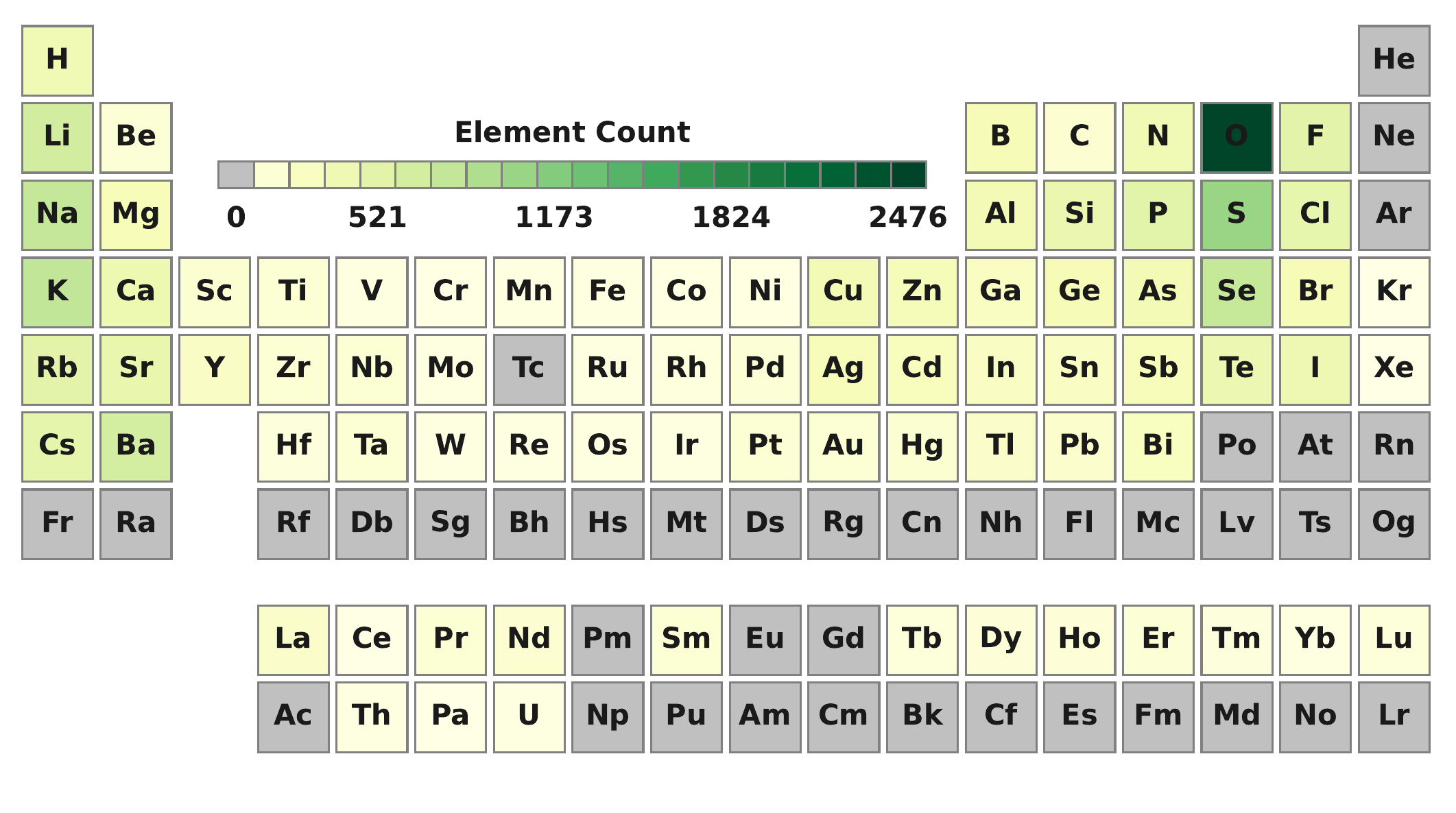}
        \caption{\textbf{Element count on the 6,674 compounds comprising the employed DFT phonons dataset.}
                The dataset spans over $79$ chemical elements of the periodic table. Most abundant atomic 
		species in the DFT phonons dataset are oxygen, sulfur, selenium, lithium, sodium and potassium.}
    \label{fig2}
\end{figure*}

\subsection{Phase-transition screening strategy}
\label{subsec:strategy}
Our high-throughput search for solid-solid phase transitions was conducted using the Materials Project database 
\cite{mp}, which contains over $170,000$ inorganic crystal structures along with their static lattice energies 
computed with DFT methods, $E_{\rm DFT}^{\rm MP}$. By restricting our selection to nonmetallic and nonmagnetic 
materials (since electronic and spin entropy contributions have been systematically disregarded throughout this
work) composed exclusively of earth-abundant elements, we reduced the pool of surveyed compounds to approximately 
$50,000$. 

Predicting temperature-induced phase transitions between different polymorphs of the same material requires knowledge 
of their vibrational free energies, which, in turn, depend on their full phonon spectra. However, performing phonon 
calculations for such a large number of structures is computationally prohibitive. To overcome this challenge, we 
developed an advanced and accurate machine-learning (ML) model to estimate vibrational free energies, $F_{\rm v}^{\rm ML}$, 
directly from atomic structure and composition (Sec.\ref{subsec:ml-fv}). This approach allowed us to approximate the 
Helmholtz free energy of each crystal structure in our dataset as (Fig.\ref{fig1}a):
\begin{equation}  
	F(V,T) \approx E_{\rm DFT}^{MP} (V) +  F_{\rm v}^{\rm ML} (V,T)~.  
\label{eq:fenerapprox}  
\end{equation}

\begin{figure*}[t]
    \centering
    \includegraphics[width=1.0\linewidth]{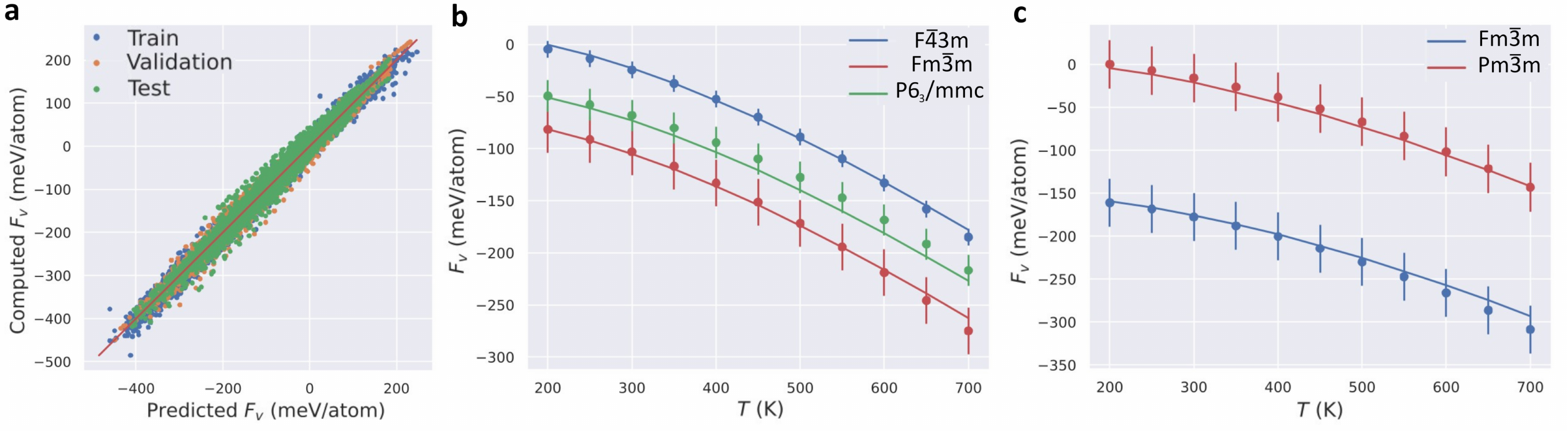}
        \caption{\textbf{Performance and validation of the developed ML vibrational free-energy model.}
         \textbf{a.}~Parity plot showing computed versus predicted vibrational free energies for the train, validation,
         and test datasets, demonstrating excellent agreement across all datasets. Temperature-dependent vibrational
         free-energy predictions (dots) for \textbf{b.}~three polymorphs of MgS and \textbf{c.}~two polymorphs of NaH,
         compared with DFT-derived $F_{\rm v}$ values (solid lines). Error bars denote uncertainty estimates based on
         latent-space graph distances.}
    \label{fig3}
\end{figure*}

\subsection{ML vibrational free-energy model}
\label{subsec:ml-fv}
Graph convolutional neural networks (GCNN) were selected to develop our ML-based vibrational free-energy model 
(Methods). Graphs, consisting of nodes (describing atoms) and edges (describing interatomic forces), are well-suited 
for representing materials due to their invariance under rotation, reflection, and translation operations 
(Fig.~\ref{fig1}b). In GCNN, convolutional layers facilitate message passing between nodes, aggregating information 
from neighboring atoms and updating node embeddings based on the graph structure. Stacking multiple convolutional 
layers enables the model to capture increasingly complex interactions and atomic dependencies in crystals \cite{gcnn1}.  

Our ML model, $F_{\rm v}^{\rm ML}$, is designed to predict the vibrational free energy of a given crystal structure 
across the temperature interval $200 \le T \le 700$~K. To achieve high accuracy, we employed a DFT-calculated dataset 
comprising $6,674$ well-behaved full phonon spectra (i.e., without imaginary frequencies) spanning a diverse range of 
materials (Fig.\ref{fig2}) \cite{togodb,hudb}. The training, validation, and testing of $F_{\rm v}^{\rm ML}$ were 
conducted using the $F_{\rm v}$ curves (Eq.~\ref{eq:fharm}) extracted from this extensive and reliable phonon dataset. 

Overall, the performance of our $F_{\rm v}^{\rm ML}$ model is very good within the whole range of considered vibrational 
free energies, which approximately extends from $-400$ to $200$~meV/atom (Fig.\ref{fig3}a). For training, validating 
and testing purposes, we used a $80$\%, $10$\% and $10$\% of the DFT phonon dataset, respectively. Our final ML model 
achieves a root mean square error of $8.23$, $11.34$, $12.31$~meV/atom for training, validation and testing, respectively. 
These values are comparable in size to the typical numerical uncertainties encountered in QH free-energy DFT calculations 
\cite{qh4}.  

For a given crystal structure, our $F_{\rm v}^{\rm ML}$ model predicts its vibrational free energy at a set of discrete 
temperature points. To obtain a continuous and differentiable vibrational free-energy function, which is crucial for 
accurately determining phase transition temperatures and deriving thermodynamic quantities such as the entropy, we fitted 
the predicted free-energy values to a simple polynomial function of the form:
\begin{equation}
    F_{\rm v}^{\rm ML} (T) \approx \alpha + \beta T^2 + \gamma T^4~,
\label{eq:smoothing}
\end{equation}
where $\alpha$, $\beta$ and $\gamma$ are real-value parameters. This polynomial smoothing function is physically motivated, 
as it ensures the correct behavior of key thermodynamic quantities, such as the vibrational entropy ($S_{\rm v} = - \partial 
F_{\rm v} / \partial T$) and heat capacity ($C_{\rm v} / T = - \partial^2 F_{\rm v} / \partial T^2$), particularly in the 
zero-temperature limit. 

Figures~\ref{fig3}b,c present validation tests for our $F_{\rm v}^{\rm ML}$ model using materials not included in the DFT 
phonon dataset considered for training. Specifically, we compare the model predictions with DFT-calculated vibrational free 
energies for three polymorphs of MgS (Fig.~\ref{fig3}b) and two polymorphs of NaH (Fig.~\ref{fig3}c). These materials were 
selected from the Materials Project database \cite{mp} based on their chemical simplicity and consistency with the presence 
of atomic species in the training dataset. As shown in the figures, the predictions of the $F_{\rm v}^{\rm ML}$ model (dots) 
exhibit excellent agreement with the DFT-calculated vibrational free energies (solid lines). Notably, the model accurately 
captures vibrational free-energy differences between polymorphs of the same compound, an essential feature for predicting 
reliable phase transitions. 

To evaluate the reliability of our $F_{\rm v}^{\rm ML}$ model in predicting vibrational free energies for materials not 
included in the original DFT phonon dataset, we incorporated uncertainty quantification (UQ) in our analysis (e.g., error 
bars in Figs.\ref{fig3}b,c). Our UQ approach assesses the similarity between a given material and those in the DFT phonon 
dataset using hyper-representation metrics, as we elaborate further below.

\subsection{Uncertainty quantification}
\label{subsec:uncertainty}
State-of-the-art uncertainty quantification (UQ) techniques primarily address the epistemic component of predictive 
uncertainty \cite{uq1}. Broadly, these approaches fall into two categories: (i)~methods that directly probe model 
architecture limitations, often paired with calibration techniques \cite{uq2}, and (ii)~methods that assess the similarity 
between new inputs and the training data.

Bayesian frameworks remain among the most established UQ strategies \cite{uq3}, incorporating probabilistic distributions 
into model parameters to estimate uncertainty. Similarly, ensemble learning methods, based on aggregating predictions from 
independently trained models or leveraging stochastic training procedures (e.g., dropout), offer robust uncertainty estimates. 
However, these approaches are often computationally prohibitive for large-scale applications.

More recently, latent space-based UQ methods have emerged as a promising alternative \cite{uq4}. These approaches assess 
uncertainty by quantifying the similarity between latent embeddings of target and training data. GCNN, in particular, are 
well-suited to this task \cite{uq5,uq6}, owing to their ability to capture complex structural and chemical relationships. 
Despite this potential, the field still lacks standardized, systematic metrics for evaluating distances in latent space, 
limiting broader adoption.

In this work, we adopt a latent distance-based UQ strategy. Specifically, we extract the hyper-representation vector 
generated by the GCNN after the pooling layer and quantify uncertainty as the square root of the Euclidean distance between 
this vector and the latent embeddings of materials in the dataset. This comparison spans the training, validation, and test 
sets, provided the model performs reliably on the latter. Our streamlined UQ approach enables efficient estimation of the 
accuracy of predicted vibrational free energies, accounting for both the chemical composition and temperature conditions 
of each prediction (Figs.\ref{fig3}b,c).

\begin{table*}[t]
    \centering
    \begin{tabular}{c c c c c}
        \hline
        \hline
         & & & & \\
        \quad Compound \qquad & \quad Phase~1 \qquad &  \quad Phase~2 \qquad & \quad $T_{t}$~(K) \qquad &
        \quad $\Delta S_{t}$~(J~K$^{-1}$~kg$^{-1}$) \qquad \\
        & & & & \\
        \hline
        & & & & \\
	    CsGaP$_{3}$HO$_{10}$ 		& C2~(P)			& P2/c~(NP) 		 & 306.9 	& 93.6  \\
	    Ta$_{3}$Bi$_{7}$O$_{18}$ 		& P1~(P) 			& C2/m~(NP) 		 & 307.3 	& 15.5  \\
	    CaZn$_{2}$(PO$_{5}$)$_{2}$ 		& Pca2$_{1}$~(P) 		& Pbcn~(NP) 		 & 309.2 	& 135.8 \\
	    ZrSeO 				& P2$_{1}$3~(P) 		& P4/nmm~(NP) 		 & 311.7 	& 70.1  \\
	    WSe$_{2}$ 				& P6$_{3}$/mmc~(NP) 		& P$\overline{6}$m2~(P)  & 312.5 	& 34.2  \\
	    Li$_{3}$TiFe$_{3}$O$_{8}$ 		& P6$_{3}$mc~(P) 		& R$\overline{3}$m~(NP)  & 313.0 	& 152.9 \\
	    K$_{2}$Ti(Si$_{2}$O$_{5}$)$_{3}$ 	& P$\overline{1}$~(NP)		& Cc~(P) 		 & 313.1 	& 26.5  \\
	    Ba$_{2}$MgCrMoO$_{6}$ 		& Immm~(NP) 			& Cm~(P) 		 & 313.5 	& 40.7  \\
	    BaZn$_{2}$Si$_{2}$O$_{7}$ 		& Cmc2$_{1}$~(P) 		& Cmcm~(NP) 		 & 314.5 	& 53.2  \\
	    Li$_{5}$Mn$_{5}$(SeO$_{3}$)$_{8}$ 	& P1~(P) 			& P$\overline{1}$~(NP) 	 & 314.7 	& 34.5  \\
	    CaAl$_{2}$SiO$_{6}$ 		& P2/c~(NP) 			& C2~(P)		 & 318.9 	& 53.6  \\
	    Ba$_{2}$SrI$_{6}$ 			& P2$_{1}$/c~(NP) 		& P$\overline{4}$b2~(P)  & 321.1 	& 18.0  \\
	    BaYNbSnO$_{6}$ 			& P$\overline{1}$~(NP) 		& P222~(P) 		 & 321.4 	& 14.9  \\
	    LiAgF$_{2}$ 			& R$\overline{3}$m~(NP) 	& C2~(P) 		 & 322.2 	& 69.0  \\
	    Li$_{3}$CoPCO$_{7}$ 		& P2$_{1}$~(P) 			& P2$_{1}$/m~(NP) 	 & 322.5 	& 59.5  \\
	    Si(PbO$_{2}$)$_{2}$ 		& C2/c~(NP) 			& C2~(P) 		 & 322.8 	& 9.8   \\
	    Y$_{2}$Zr$_{2}$O$_{7}$ 		& Fd$\overline{3}$m~(NP) 	& P2$_{1}$~(P) 		 & 323.6 	& 18.8  \\
	    LiGa(H$_{2}$N)$_{4}$ 		& P2$_{1}$~(P) 			& P2$_{1}$/c~(NP) 	 & 324.0 	& 50.9  \\
	    LiBi$_{3}$(IO$_{2}$)$_{2}$ 		& Cmcm~(NP) 			& Amm2~(P) 		 & 324.2 	& 17.4  \\
	    LiBiB$_{2}$O$_{5}$ 			& C2/c~(NP) 			& C2~(P) 		 & 325.1 	& 50.4  \\
	    Mn$_{3}$Zn$_{2}$O$_{8}$ 		& P6$_{3}$mc~(P) 		& C2/m~(NP) 		 & 326.0 	& 85.9  \\
	    Li$_{3}$Cr(CoO$_{3}$)$_{2}$ 	& P$\overline{1}$~(NP) 		& P1~(P) 		 & 328.5 	& 57.5  \\
	    LiAl(Si$_{2}$O$_{5}$)$_{2}$ 	& Pc~(P) 			& P2/c~(NP) 	 	 & 332.6 	& 51.9  \\
	    Cs$_{3}$Sb$_{2}$Cl$_{9}$ 		& P$\overline{3}$m1~(NP) 	& P321~(P) 		 & 333.0 	& 7.5   \\
	    MoWSe$_{4}$ 			& P$\overline{3}$m1~(NP) 	& P3m1~(P) 		 & 333.3 	& 5.0   \\
	    VBi(PbO$_{3}$)$_{2}$ 		& P2$_{1}$/c~(NP) 		& Pc~(P) 		 & 334.4 	& 9.1   \\
	    Ba$_{5}$P$_{3}$O$_{12}$F 		& P6$_{3}$/m~(NP) 		& P6$_{3}$~(P) 		 & 335.8 	& 17.4  \\
	    ErCu(WO$_{4}$)$_{2}$ 		& P$\overline{1}$~(NP) 		& P1~(P) 		 & 337.3 	& 22.8  \\
	    LiCr$_{3}$O$_{8}$ 			& Pmn2$_{1}$~(P) 		& C2/m~(NP) 		 & 337.5 	& 58.0  \\
	    MgFe$_{4}$(PO$_{4}$)$_{4}$ 		& Pm~(P) 			& P$\overline{1}$~(NP) 	 & 337.9 	& 33.3  \\
	    Li$_{2}$V$_{2}$O$_{5}$F$_{2}$ 	& P$\overline{1}$~(NP) 		& Cc~(P) 		 & 338.2 	& 117.0 \\
	    Hg$_{3}$(SCl)$_{2}$ 		& I2$_{1}$3~(P) 		& Pm$\overline{3}$n~(NP) & 339.1 	& 30.6  \\
	    TbCuTe$_{2}$ 			& P3m1~(P) 			& P$\overline{3}$m1~(NP) & 339.1 	& 25.3  \\
	    P$_{2}$SN$_{3}$Cl$_{5}$O 		& P$\overline{1}$~(NP) 		& P2$_{1}$2$_{1}$2$_{1}$~(P) & 339.8 	& 139.2 \\
	    NaVP$_{2}$O$_{7}$ 			& P2$_{1}$~(P) 			& P2$_{1}$/c~(NP)        & 340.0 	& 8.1   \\
	    K$_{5}$HfIn(MoO$_{4}$)$_{6}$ 	& R3~(P) 			& R$\overline{3}$c~(NP)  & 340.6 	& 12.6  \\
	    KSb$_{2}$PO$_{8}$ 			& Cc~(P) 			& C2/c~(NP) 		 & 342.3 	& 18.1  \\
	    Mn$_{3}$(OF$_{3}$)$_{2}$ 		& Pc~(P) 			& P$\overline{1}$~(NP) 	 & 344.8 	& 72.4  \\
	    Li$_{8}$(CoO$_{2}$)$_{5}$ 		& P1~(P) 			& P$\overline{1}$~(NP) 	 & 346.0 	& 40.5  \\
	    Li$_{2}$MnF$_{6}$ 			& P4$_{2}$/mnm~(NP) 		& P321~(P) 		 & 346.2 	& 145.4 \\
	    NaLi$_{2}$FePCO$_{7}$ 		& P$\overline{1}$~(NP) 		& P1~(P) 		 & 346.4 	& 84.2  \\
	    Bi$_{2}$W$_{2}$O$_{9}$ 		& Pna2$_{1}$~(P) 		& Pbcn~(NP) 		 & 346.5 	& 9.3   \\
	    PNCl$_{2}$ 				& P2$_{1}$2$_{1}$2$_{1}$~(P) 	& Pnma~(NP) 		 & 348.5 	& 338.3 \\
	    K$_{2}$ZnBr$_{4}$ 			& P2$_{1}$/m~(NP) 		& P2$_{1}$~(P) 		 & 349.8 	& 6.8   \\
	& & & & \\
        \hline
        \hline
    \end{tabular}
	\caption{\textbf{Stable polar-nonpolar phase transitions predicted to occur in the temperature interval $300 \le 
	T_{t} \le 350$~K.} The low-$T$ polymorph corresponds to ``Phase 1'' and the high-$T$ polymorph to ``Phase 2''. 
	``P'' and ``NP'' stand for polar and nonpolar structures, respectively. $T_{t}$ and $\Delta S_{t}$ stand for 
	phase-transition temperature and entropy change, respectively.}
    \label{tab1a}
\end{table*}

\begin{table*}[t]
    \centering
    \begin{tabular}{c c c c c}
        \hline
        \hline
         & & & & \\
        \quad Compound \qquad & \quad Phase~1 \qquad &  \quad Phase~2 \qquad & \quad $T_{t}$~(K) \qquad &
        \quad $\Delta S_{t}$~(J~K$^{-1}$~kg$^{-1}$) \qquad \\
        & & & & \\
        \hline
        & & & & \\
	    Fe$_{4}$OF$_{7}$		 		& P1 				& Cmc2$_{1}$ 			& 300.4 & 252.1 \\
	    Mo(WS$_{3}$)$_{2}$		 		& P3m1 				& P$\overline{6}$m2 		& 302.2 & 8.9 \\
	    DyCuS$_{2}$			 		& P2$_{1}$2$_{1}$2$_{1}$ 	& P2$_{1}$2$_{1}$2$_{1}$ 	& 302.2 & 5.2 \\
	    Li$_{3}$Fe(PO$_{4}$)$_{2}$   		& P2$_{1}$ 			& C2 				& 305.7 & 53.1 \\
	    SbCl$_{3}$F$_{2}$		 		& I4 				& I$\overline{4}$		& 307.4 & 8.9 \\
	    WN$_{2}$			 		& P3$_{1}$21 			& Pna2$_{1}$ 			& 309.3 & 15.0 \\
	    MgTe 			 		& P6$_{3}$mc 			& F$\overline{4}$3m		& 309.9 & 25.6 \\
	    VOF$_{3}$ 					& P2$_{1}$2$_{1}$2$_{1}$ 	& Pc 				& 310.6 & 241.7 \\
	    MnTeMoO$_{6}$ 				& P2$_{1}$2$_{1}$2   		& P$\overline{4}$2$_{1}$m	& 312.5 & 64.5 \\
	    Nb$_{4}$Ag$_{2}$O$_{11}$      		& R3 				& R3c 				& 315.1 & 58.4 \\
	    Li$_{3}$V(H$_{4}$O$_{3}$)$_{4}$		& P1 				& Pc 				& 316.1 & 122.8 \\
	    Al$_{3}$CrO$_{6}$				& R3 				& Pc 				& 318.3 & 154.2 \\
	    LiMoP$_{4}$O$_{13}$				& P1 				& P2$_{1}$2$_{1}$2$_{1}$ 	& 319.8 & 56.0 \\
	    KNb(BO$_{3}$)$_{2}$ 			& Pna2$_{1}$ 			& Pmn2$_{1}$ 			& 321.6 & 195.5 \\
	    Fe$_{2}$P$_{2}$O$_{7}$ 			& P1 				& P1 				& 322.3 & 12.5 \\
	    Li$_{2}$TiFeO$_{4}$ 			& P1 				& P1 				& 324.3 & 16.9 \\
	    CO 						& P2$_{1}$2$_{1}$2$_{1}$ 	& R3c 				& 324.9 & 161.9 \\
	    K$_{4}$BaTi$_{6}$S$_{20}$O 			& P6$_{3}$22 			& Fdd2 				& 325.3 & 34.9 \\
	    MnInF$_{3}$					& P1 				& R3m 				& 325.7 & 12.9 \\
	    LiMn$_{3}$OF$_{5}$ 				& P1 				& Cc 				& 326.3 & 31.3 \\
	    Li$_{2}$V$_{3}$WO$_{8}$ 			& P6$_{3}$mc 			& P1 				& 329.5 & 82.2 \\
	    La$_{4}$FeSe$_{6}$O				& P1 				& P1 				& 329.8 & 5.0 \\
	    Ta$_{7}$BiO$_{19}$ 				& P$\overline{6}$c2 		& P6$_{3}$22 			& 330.4 & 54.7 \\
	    K$_{2}$Al$_{2}$B$_{2}$O$_{7}$ 		& C2 				& C2 				& 332.1 & 41.8 \\
	    K$_{2}$Zn$_{6}$O$_{7}$ 			& P4$_{2}$nm 			& P4$_{2}$nm 			& 332.3 & 34.2 \\
	    Li$_{6}$Mn$_{5}$Fe(BO$_{3}$)$_{6}$  	& P1 				& Cm 				& 332.8 & 56.8 \\
	    Na$_{5}$Zr$_{4}$Si$_{3}$(PO$_{8}$)$_{3}$ 	& P1 				& R32 				& 333.2 & 35.4 \\
	    CsGeI$_{3}$ 				& Cm 				& Cm 				& 337.8 & 55.4 \\
	    Sr$_{2}$MgZrCrO$_{6}$ 			& P1 				& P1 				& 339.1 & 5.1 \\
	    Li$_{4}$MnV$_{3}$(P$_{2}$O$_{7}$)$_{4}$ 	& P1 				& P2$_{1}$ 			& 339.6 & 83.5 \\
	    Li$_{5}$Cu(PO$_{4}$)$_{2}$ 			& P1 				& Pm				& 341.1 & 103.1 \\
	    ZnTe 					& F$\overline{4}$3m    		& P6$_{3}$mc 			& 342.1 & 5.4 \\
	    Ge$_{3}$Pb$_{5}$O$_{11}$ 			& P3 				& P$\overline{6}$ 		& 343.2 & 7.4 \\
	    Mg$_{2}$TiO$_{4}$ 				& P4$_{1}$22 			& P1 				& 344.4 & 209.5 \\
	    HoFe$_{3}$(BO$_{3}$)$_{4}$ 			& P3$_{1}$21 			& R32 				& 344.8 & 68.9 \\
	    Li$_{3}$MnF$_{7}$ 				& R3m 				& P1 				& 346.6 & 342.3 \\
	    MgFeF$_{4}$ 				& P1 				& P1 				& 347.1 & 10.9 \\
	    MgFe$_{3}$Ni$_{2}$Sb(PO$_{4}$)$_{6}$ 	& P1 				& P1 				& 348.4 & 36.8 \\
        & & & & \\
        \hline
        \hline
    \end{tabular}
        \caption{\textbf{Stable polar-polar phase transitions predicted to occur in the temperature interval $300 \le 
	T_{t} \le 350$~K.} The low-$T$ polymorph corresponds to ``Phase 1'' and the high-$T$ polymorph to ``Phase 2''. 
	$T_{t}$ and $\Delta S_{t}$ stand for phase-transition temperature and entropy change, respectively.}
    \label{tab1b}
\end{table*}

\begin{table*}[t]
    \centering
    \begin{tabular}{c c c c c}
        \hline
        \hline
         & & & & \\
        \quad Compound \qquad & \quad Phase~1 \qquad &  \quad Phase~2 \qquad & \quad $T_{t}$~(K) \qquad &
        \quad $\Delta S_{t}$~(J~K$^{-1}$~kg$^{-1}$) \qquad \\
        & & & & \\
        \hline
        & & & & \\
	    CdSeO$_{3}$ 			& Pnma 				& P2$_{1}$/c 			& 301.6 & 58.6 \\
	    NaMnO$_{4}$ 			& Cmcm 				& P2$_{1}$/c 			& 302.5 & 41.1 \\
	    Sr$_{2}$SnO$_{4}$ 			& Pccn 				& Cmce 				& 304.2 & 1.5 \\
	    CaPbI$_{4}$ 			& C2/m 				& P2/m 				& 304.8 & 1.7 \\
	    K$_{2}$FeF$_{5}$ 			& C2/m 				& Pbam 				& 305.0 & 31.3 \\
	    Na$_{4}$TiP$_{2}$O$_{9}$ 		& Cmme 				& P2/c 				& 305.8 & 25.0 \\
	    Ba(CuS)$_{2}$ 			& Pnma 				& I4/mmm 			& 306.8 & 77.6 \\
	    KLaTiO$_{4}$ 			& Pbcm 				& P4/nmm 			& 306.9 & 23.8 \\
	    Na$_{2}$Ti$_{2}$O$_{5}$ 		& Pbcn 				& C2/m 				& 309.3 & 36.5 \\
	    Sr$_{2}$ZnWO$_{6}$ 			& P$\overline{1}$ 		& P2$_{1}$/c 			& 309.4 & 30.6 \\
	    LiLa$_{4}$MnO$_{8}$ 		& I4$_{1}$/amd 			& Cmmm 				& 311.4 & 2.3 \\
	    Na$_{2}$FePCO$_{7}$ 		& P2$_{1}$/m 			& P$\overline{1}$ 		& 312.0 & 137.1 \\
	    ZnCoO$_{2}$				& R$\overline{3}$m 		& C2/m 				& 314.0 & 127.9 \\
	    CdPS$_{3}$ 				& C2/m 				& R$\overline{3}$ 		& 314.2 & 28.7 \\
	    BaYBr$_{5}$ 			& P2$_{1}$/c 			& C2/c 				& 314.5 & 20.2 \\
	    Zn$_{2}$Ga$_{2}$S$_{5}$ 		& R$\overline{3}$m 		& P6$_{3}$/mmc 			& 315.0 & 80.5 \\
	    NdBO$_{3}$ 				& Pnma 				& P$\overline{1}$ 		& 316.6 & 132.5 \\
	    Li$_{6}$SbS$_{2}$ 			& C2/c 				& P2$_{1}$/c 			& 316.7 & 41.6 \\
	    Nd(BO$_{2}$)$_{3}$ 			& Pnma 				& C2/c 				& 317.1 & 62.1 \\
	    Li$_{2}$Cr$_{2}$(SO$_{4}$)$_{3}$ 	& Pbcn 				& C2/c 				& 318.3 & 142.6 \\
	    Li$_{5}$AlO$_{4}$ 			& Pbca 				& Pmmn 				& 319.0 & 75.6 \\
	    K$_{2}$Mg(CO$_{3}$)$_{2}$ 		& R$\overline{3}$m 		& C2/m 				& 319.7 & 129.0 \\
	    Li$_{2}$Sn$_{2}$(SO$_{4}$)$_{3}$ 	& Pbca 				& Pbcn 				& 322.1 & 24.5 \\
	    SrC$_{2}$ 				& C2/c 				& I4/mmm 			& 322.4 & 123.4 \\
	    WCl$_{6}$ 				& P$\overline{3}$m1		& R$\overline{3}$ 		& 322.8 & 8.9 \\
	    SrCN$_{2}$ 				& R$\overline{3}$m 		& Pnma 				& 323.4 & 110.0 \\
	    LiFe$_{3}$O$_{4}$ 			& P$\overline{1}$ 		& C2/c 				& 324.8 & 139.7 \\
	    VBiO$_{4}$ 				& I4$_{1}$/a 			& I4$_{1}$/amd 			& 327.0 & 33.9 \\
	    W$_{5}$(O$_{2}$F$_{11}$)$_{2}$ 	& P$\overline{1}$ 		& I4$_{1}$/a 			& 327.2 & 12.9 \\
	    Cs$_{2}$Se 				& Pnma 				& R$\overline{3}$m 		& 327.3 & 113.5 \\	
	    Sn$_{2}$N$_{2}$O 			& I4$_{1}$/amd 			& P$\overline{3}$m1 		& 331.6 & 12.3 \\
	    SeN 				& P2$_{1}$/c 			& C2/c 				& 331.8 & 73.0 \\
	    MgP$_{2}$(H$_{8}$O$_{5}$)$_{2}$ 	& I4$_{1}$/acd 			& P4$_{2}$/nmc 			& 332.7 & 44.7 \\
	    Ba$_{2}$SmNbO$_{6}$ 		& Fm$\overline{3}$m 		& Pn$\overline{3}$ 		& 333.2 & 5.7 \\
	    Na$_{2}$Al$_{2}$B$_{2}$O$_{7}$ 	& P2$_{1}$/m 			& P$\overline{3}$1c 		& 333.3 & 52.3 \\
	    Mg$_{2}$PHO$_{5}$ 			& Pnma 				& P2$_{1}$/c 			& 334.1 & 86.8 \\
	    V$_{2}$FeO$_{6}$ 			& P$\overline{1}$ 		& C2/c 				& 334.3 & 42.9 \\
	    ZnCl$_{2}$ 				& P4$_{2}$/nmc 			& P2$_{1}$/c 			& 334.3 & 4.2 \\
	    Li$_{3}$Sn$_{2}$(PO$_{4}$)$_{3}$ 	& P2$_{1}$/c 			& R$\overline{3}$ 		& 334.7 & 149.3 \\
	    SF$_{6}$ 				& C2/m 				& Im$\overline{3}$m 		& 340.4 & 130.8 \\
	    KAlF$_{4}$ 				& P4/mbm 			& P4/mmm 			& 342.4 & 55.0 \\
	    CaSeO$_{4}$ 			& I4$_{1}$/a 			& Cmce 				& 344.4 & 21.6 \\
	    P$_{3}$(WO$_{6}$)$_{2}$ 		& R$\overline{3}$ 		& P2$_{1}$/c 			& 344.8 & 57.0 \\
	    ErBiW$_{2}$O$_{9}$ 			& Pnma 				& Pmmn 				& 345.6 & 13.8 \\
	    Ca(CoS$_{2}$)$_{2}$ 		& Fd$\overline{3}$m 		& I4$_{1}$/amd 			& 345.8 & 27.8 \\
	    LiMn$_{2}$F$_{5}$ 			& Cmcm 				& Pnnm 				& 349.8 & 31.0 \\
	    Ba$_{2}$SmVO$_{6}$ 			& Fm$\overline{3}$m 		& Pn$\overline{3}$ 		& 349.9 & 38.5 \\
	    Sr$_{2}$Bi$_{2}$Se$_{3}$O$_{2}$     & P2$_{1}$/c 			& C2/m 				& 350.0 & 10.8 \\
        & & & & \\
        \hline
        \hline
    \end{tabular}
        \caption{\textbf{Stable nonpolar-nonpolar phase transitions predicted to occur in the temperature interval $300 
	\le T_{t} \le 350$~K.} The low-$T$ polymorph corresponds to ``Phase 1'' and the high-$T$ polymorph to ``Phase 2''. 
        $T_{t}$ and $\Delta S_{t}$ stand for phase-transition temperature and entropy change, respectively.}
    \label{tab1c}
\end{table*}

\begin{figure*}[t]
    \centering
    \includegraphics[width=1.0\linewidth]{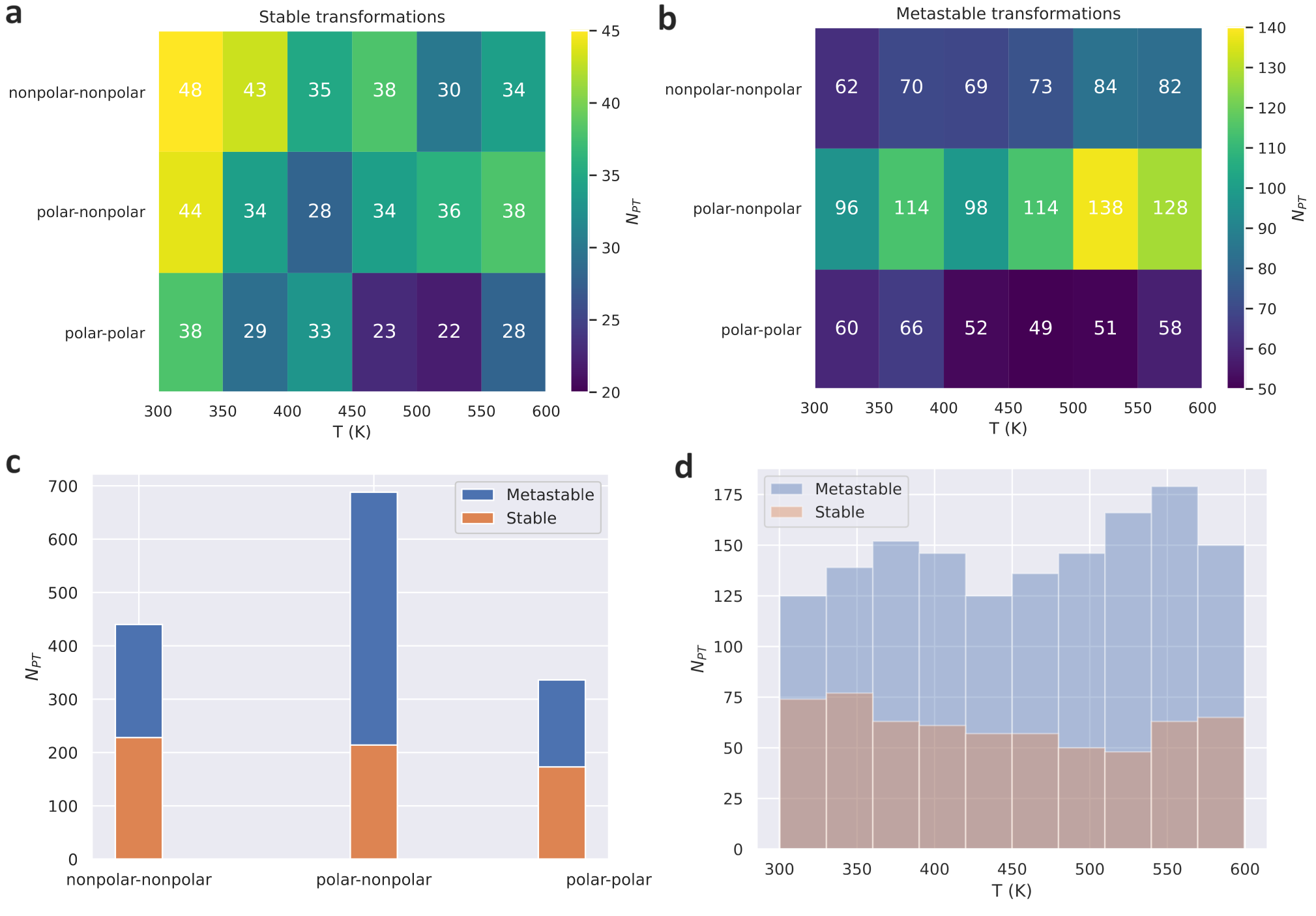}
	\caption{\textbf{Number of predicted solid-solid phase transformations, $N_{PT}$, in the temperature interval 
	$300 \le T_{t} \le 600$~K expressed as a function of thermodynamic stability, transition temperature and crystal 
	symmetry properties.} \textbf{a.}~Number of predicted stable and \textbf{b.}~metastable phase transitions. 
	\textbf{c.}~Number of predicted phase transitions as a function of stability and symmetry. 
	\textbf{d.}~Number of predicted phase transitions as a function of stability and temperature.
        }
    \label{fig4}
\end{figure*}

\begin{figure*}[t]
    \centering
    \includegraphics[width=1.0\linewidth]{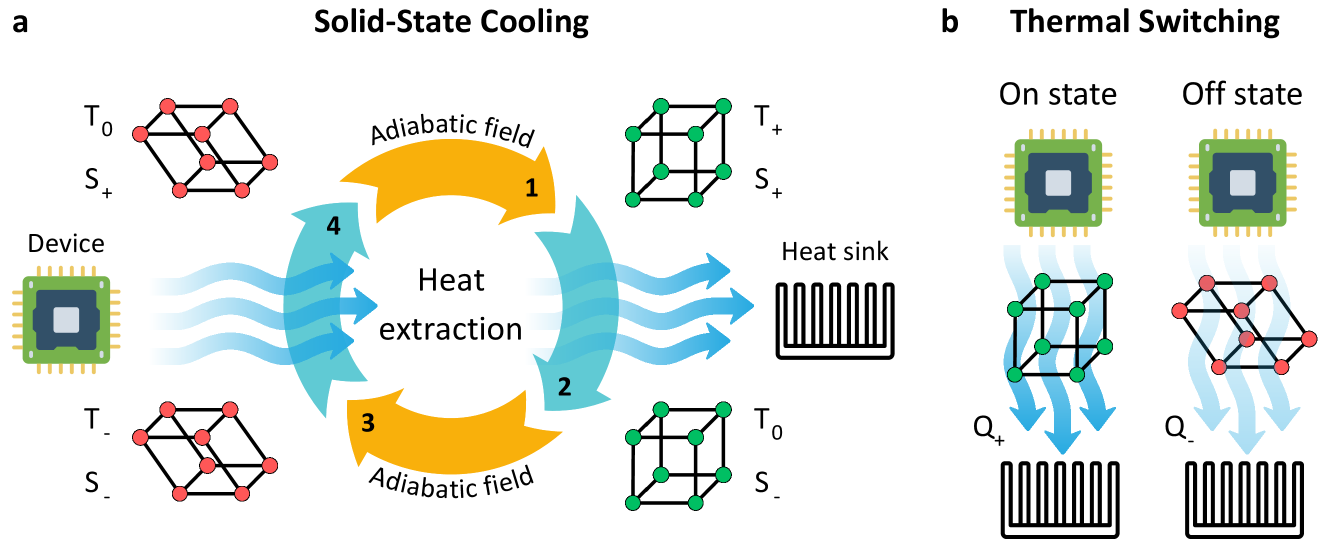}
        \caption{\textbf{Sketch of two thermal management applications in which our solid-solid phase transition
	predictions may be impactful.}
	\textbf{a.}~Solid-state cooling, an energy efficient and sustainable technology that employs crystals instead of 
	enviromentally harmful gases to refrigerate.   
	\textbf{b.}~Thermal switching, where large thermal conductivity changes are engineered for developing novel 
	approaches to energy harvesting and neuromorphic and phononic computing.} 
    \label{fig5}
\end{figure*}

\section{Results}
\label{sec:results}
We begin this section by presenting the general results of our ML-guided high-throughput screening for $T$-induced 
solid-solid phase transitions in inorganic crystals. Our study confidently predicts a total of $2,079$ polymorphic transformations 
in the temperature interval $300 \le T \le 600$~K, which are detailed in the Supplementary Material. To ensure reliability, 
we discarded phase transitions where vibrational free-energy uncertainties exceeded $50$\% of the difference 
$|\Delta F_{\rm v}^{\rm ML}| \equiv |F_{\rm v}^{\rm ML} (600~\rm{K}) - F_{\rm v}^{\rm ML} (300~\rm{K})|$ (Sec.\ref{subsec:uncertainty}), 
as their transition temperatures could not be assessed with confidence. We then highlight two key thermal management applications 
where our findings could have a significant impact: solid-state cooling and thermal conductivity switching.

\subsection{$T$-induced solid-solid phase transitions}
\label{subsec:general}
We predicted $615$ stable phase transitions (Fig.\ref{fig1}c and Supplementary Material), with $130$ occurring within 
the technologically relevant temperature interval $300 \le T_{t} \le 350$~K (Fig.\ref{fig4}a). Metastable phase transitions 
(Fig.\ref{fig1}c) are also of interest, as they can be induced not only by temperature but also by external fields 
such as pressure or electric bias. Our study identified $1,464$ metastable phase transitions (Supplementary Material), 
including $218$ within the $300$--$350$~K temperature range (Fig.\ref{fig4}b).  

In this study, we classify the predicted $T$-induced phase transitions according to their thermodynamic stability, 
transition temperature, and crystal symmetry properties (Figs.\ref{fig4}c,d). Regarding crystal symmetry, we classify 
phases as ``nonpolar'' if they are centrosymmetric and ``polar'' if they are noncentrosymmetric (e.g., enantiomorphic). 
Accordingly, phase transitions fall into three main types: ``polar-polar'', ``polar-nonpolar'', and ``nonpolar-nonpolar''. 
From a functional perspective, polar-nonpolar and polar-polar transitions are particularly interesting (Tables~1 and 2), 
as they can be induced by external electric fields, similar to ferroelectrics \cite{davis06,cazorla18,moriwake16}.  

Knowing the latent heat, $L$, of a phase transition is important both from a fundamental perspective and for practical 
applications. $L$ measures the energy required to induce a phase change in a material without altering its temperature 
and is directly related to the entropy change of the phase transition via $\Delta S_{t} \equiv L / T_{t}$. In materials 
science, $L$ is crucial for assessing thermal storage capacity, heat management strategies, and the suitability of materials 
for energy-related phase change applications. For example, materials with large $L$ can be employed in thermal 
energy storage systems, where energy is absorbed or released during phase transitions. Furthermore, knowledge of $L$ 
enables classification of the phase transition into first-order (discontinuous, $L \neq 0$) or second-order (continuous, 
$L \approx 0$). For these reasons, we report the entropy change $\Delta S_{t}$ for most phase transitions predicted in 
this study (Tables~1--5 and Supplementary Material).

For stable phase transitions, we predict $228$ nonpolar-nonpolar, $214$ polar-nonpolar, and $173$ polar-polar transitions 
occurring within the $300 \le T_{t} \le 600$~K range (Figs.\ref{fig4}a,c). For metastable phase transitions, we identify $440$ 
nonpolar-nonpolar, $688$ polar-nonpolar, and $336$ polar-polar transitions in the same temperature range (Figs.\ref{fig4}b,c).  
In all three categories, metastable transitions outnumber the stable ones. This trend can be attributed to the fact that 
considering multiple metastable polymorphs increases the number of possible free-energy curve crossings, in contrast to 
cases where only the most stable phase and a single low-energy metastable structure are included. 

Interestingly, phase transitions involving polar phases are significantly more numerous than purely nonpolar-nonpolar ones, in 
both the stable and metastable sets (Fig.\ref{fig4}c). For example, within the temperature interval $300 \le T_{t} \le 350$~K, 
we predict a total of $92$ stable and $158$ metastable transitions involving polar phases, compared to only $38$ stable and $60$ 
metastable transitions that are purely nonpolar-nonpolar. Furthermore, we observe a slight decrease in the number of predicted 
stable transitions with increasing temperature, whereas metastable transitions exhibit the opposite trend, becoming more frequent 
as temperature rises (Fig.~\ref{fig4}d). All $2,079$ polymorphic transformations predicted within the $300 \le T \le 600$~K range, 
both stable and metastable, are detailed in the Supplementary Material. 

Tables~1 and 2 list the stable polar-nonpolar and polar-polar phase transitions occurring in the temperature interval $300 
\le T_{t} \le 350$~K, which could in practice be triggered by an external electric bias. Several key observations emerge 
from these datasets. First, complex oxide compounds dominate, accounting for approximately $74$\% of these transitions. 
This result aligns with the fact that most known room-temperature polar materials are oxides (e.g., BaTiO$_{3}$ and BiFeO$_{3}$). 
And second, many materials in Tables~1 and 2 would typically be overlooked in routine first-principles QHA studies due to their 
complex compositions (e.g., Na$_{5}$Zr$_{4}$Si$_{3}$(PO$_{8}$)$_{3}$ and Ba$_{2}$MgCrMoO$_{6}$) and low-symmetry crystal 
structures (e.g., $P1$ and $P\overline{1}$), which significantly increase computational costs. 

Nevertheless, the materials and phase transitions listed in Tables~1 and 2 are experimentally worth exploring for several 
reasons. Notably, many of the transitions reported in Table~1 involve low-$T$ nonpolar phases transforming into high-$T$ 
polar phases, a reversal of the typical behavior observed in archetypal ferroelectrics, where the high-$T$ phase is usually 
nonpolar. Moreover, some of the entropy changes associated with polar-nonpolar and polar-polar phase transitions are 
remarkably large (e.g., $\Delta S_{t} = 342.3$~J~K$^{-1}$~kg$^{-1}$ for Li$_{3}$MnF$_{7}$, $\Delta S_{t} = 
338.3$~J~K$^{-1}$~kg$^{-1}$ for PnCl$_{2}$ and $\Delta S_{t} = 252.1$~J~K$^{-1}$~kg$^{-1}$ for Fe$_{4}$OF$_{7}$), 
indicating a pronounced first-order character and suggesting potential for electrically controlled thermal management 
applications. 

Table~3 summarizes the stable nonpolar-nonpolar phase transitions predicted to occur within the temperature range  
$300$--$350$~K. As in previous cases, the reported transitions predominantly involve chemically complex compounds with 
low crystalline symmetries (e.g., Sr$_{2}$ZnWO$_{6}$ and Na$_{2}$FePCO$_{7}$). Interestingly, the entropy changes 
associated with these near room-temperature phase transitions tend to be smaller than those involving polar phases. 
For instance, the maximum $\Delta S_{t}$ reported in Table~3 is $149.3$~J~K$^{-1}$~kg$^{-1}$, compared to 
$338.3$~J~K$^{-1}$~kg$^{-1}$ for polar-nonpolar phase transitions (Table~1) and $342.3$~J~K$^{-1}$~kg$^{-1}$ for 
polar-polar transformations (Table~2). We tentatively attribute this general trend to the typically higher symmetry 
of centrosymmetric (nonpolar) phases, which may limit the magnitude of the structural and entropic changes involved 
in such phase transitions.

The large number of phase transitions predicted in this study spans a broad range of chemical compositions and crystal 
structures. Validation tests supporting these predictions will be presented in the Discussion section. To maintain focus 
and offer illustrative insights, we next highlight two key thermal management applications where the outcomes of our 
ML-guided high-throughput screening could have a significant impact (Fig.~\ref{fig5}).

\begin{table*}[h]
    \centering
    \begin{tabular}{c c c c c}
        \hline
        \hline
	& & & & \\
        \quad Compound \qquad & \quad Phase~1 \qquad &  \quad Phase~2 \qquad & \quad $T_{t}$~(K) \qquad & 
	\quad $\Delta S_{t}$~(J~K$^{-1}$~kg$^{-1}$) \qquad \\
	& & & & \\
        \hline
	& & & & \\
	    Li$_{3}$MnF$_{7}$	      		&         R3m 			&          P1     &              346.6 &          342.3 \\
	    PNCl$_{2}$		      		&  P2$_{1}$2$_{1}$2$_{1}$ 	&        Pnma 	  &              348.5 &          338.3 \\
	    Fe$_{4}$OF$_{7}$	      		&          P1 			&      Cmc2$_{1}$ &              300.4 &          252.1 \\
	    VOF$_{3}$		      		&  P2$_{1}$2$_{1}$2$_{1}$ 	&          Pc     &              310.6 &          241.7 \\
	    Mg$_{2}$TiO$_{4}$         		&      P4$_{1}$22 		&          P1     &              344.4 &          209.5 \\
	    KNb(BO$_{3}$)$_{2}$       		&      Pna2$_{1}$ 		&      Pmn2$_{1}$ &              321.5 &          195.5 \\
	    CO                        		&  P2$_{1}$2$_{1}$2$_{1}$ 	&         R3c &                  324.9 &          161.9 \\
	    Al$_{3}$CrO$_{6}$         		&          R3 			&          Pc &                  318.3 &          154.2 \\
	    Li$_{3}$TiFe$_{3}$O$_{8}$ 		&      P6$_{3}$mc 		& R$\overline{3}$m &             313.0 &          152.9 \\
	    Li$_{3}$Sn$_{2}$(PO$_{4}$)$_{3}$ 	&      P2$_{1}$/c 		& R$\overline{3}$ &              334.8 &          149.3 \\
	    Li$_{2}$MnF$_{6}$			&    P4$_{2}$/mnm 		&        P321 &                  346.2 &          145.4 \\
	    Li$_{2}$Cr$_{2}$(SO$_{4}$)$_{3}$    &        Pbcn 			&        C2/c &                  318.3 &          142.6 \\
	    LiFe$_{3}$O$_{4}$                   & P$\overline{1}$ 	        &        C2/c &                  324.8 &          139.7 \\
	    P$_{2}$SN$_{3}$Cl$_{5}$O            & P$\overline{1}$ 		&  P2$_{1}$2$_{1}$2$_{1}$ &      339.8 &          139.2 \\
	    Na$_{2}$FePCO$_{7}$                 &      P2$_{1}$/m 		&  P$\overline{1}$ &             312.0 &          137.1 \\
	    CaZn$_{2}$(PO$_{5}$)$_{2}$          &      Pca2$_{1}$ 		&        Pbcn &                  309.2 &          135.8 \\
	    NdBO$_{3}$                          &        Pnma 			&  P$\overline{1}$ &             316.6 &          132.5 \\
	    SF$_{6}$                            &        C2/m 			&  Im$\overline{3}$m &           340.4 &          130.8 \\
	    K$_{2}$Mg(CO$_{3}$)$_{2}$           & R$\overline{3}$m 		&        C2/m &                  319.7 &          129.0 \\
	    ZnCoO$_{2}$                         & R$\overline{3}$m 	        &        C2/m &                  314.0 &          127.9 \\
	    SrC$_{2}$                           &        C2/c 			&      I4/mmm &                  322.4 &          123.4 \\
	    Li$_{3}$V(H$_{4}$O$_{3}$)$_{4}$     &          P1 			&          Pc &                  316.1 &          122.8 \\
	    Li$_{2}$V$_{2}$O$_{5}$F$_{2}$       & P$\overline{1}$ 		&          Cc &                  338.2 &          117.0 \\
	    Cs$_{2}$Se                          &        Pnma 			&  R$\overline{3}$m &            327.3 &          113.5 \\
            SrCN$_{2}$                          & R$\overline{3}$m 		&        Pnma &                  323.4 &          110.0 \\
            Li$_{5}$Cu(PO$_{4}$)$_{2}$          &          P1 			&          Pm &                  341.1 &          103.1 \\
	    & & & & \\
        \hline
        \hline
    \end{tabular}
	\caption{\textbf{Stable phase transformations predicted to occur in the temperature interval $300 \le T_{t} \le 350$~K
	with a phase transition entropy change higher than $100$~J~K$^{-1}$~kg$^{-1}$.} The low-$T$ polymorph corresponds 
	to ``Phase 1'' and the high-$T$ polymorph to ``Phase 2''. $T_{t}$ and $\Delta S_{t}$ stand for phase-transition 
	temperature and entropy change, respectively. 
	    }
    \label{tab2}
\end{table*}

\begin{table*}[h]
    \centering
    \begin{tabular}{c c c c c}
        \hline
        \hline
        & & & & \\
        \quad Compound \qquad & \quad Phase~1 \qquad &  \quad Phase~2 \qquad & \quad $T_{t}$~(K) \qquad &
        \quad $\Delta S_{t}$~(J~K$^{-1}$~kg$^{-1}$) \qquad \\
        & & & & \\
        \hline
        & & & & \\
	    PNF$_{2}$                      & Pnma~(NP)                    & Pna2$_{1}$~(P)         & 410.8 & 360.7 \\
	    KNO$_{3}$                      & R3m~(P)                      & P2$_{1}$/c~(NP)        & 539.3 & 346.4 \\
	    PNCl$_{2}$                     & P2$_{1}$2$_{1}$2$_{1}$~(P)   & Pnma~(NP)              & 348.5 & 338.3 \\
	    Li$_{2}$FeF$_{5}$              & C2/c~(NP)                    & P1~(P)                 & 399.6 & 304.4 \\
	    NaClO$_{4}$                    & Pnma~(NP)                    & I$\overline{4}$2m~(P)  & 538.3 & 294.8 \\
	    LiP$_{2}$WO$_{7}$              & P2$_{1}$~(P)                 & P2$_{1}$/c~(NP)        & 593.1 & 262.8 \\
	    Cd(C$_{4}$N$_{3}$)$_{2}$       & P31m~(P)                     & Pmna~(NP)              & 369.9 & 259.0 \\
	    Li$_{2}$NiSnO$_{4}$            & I$\overline{4}$m2~(P)        & C2/c~(NP)              & 442.9 & 257.6 \\
	    Li$_{2}$VH$_{2}$OF$_{5}$       & C2/c~(NP)                    & P1~(P)                 & 568.3 & 251.4 \\
	    H$_{2}$CBrCl                   & P2$_{1}$2$_{1}$2~(P)         & C2/c~(NP)              & 558.1 & 247.8 \\
	    KCSN                           & Pbcm~(NP)                    & Ima2~(P)               & 567.0 & 247.5 \\
	    CsClO$_{4}$                    & Pnma~(NP)                    & F$\overline{4}$3m~(P)  & 596.6 & 235.0 \\
	    VF$_{4}$                       & Pc~(P)                       & C2/c~(NP)              & 486.5 & 233.8 \\
	    LiCu(PO$_{3}$)$_{3}$           & P2$_{1}$2$_{1}$2$_{1}$~(P)   & P$\overline{1}$~(NP)   & 576.9 & 232.4 \\
	    LiCuS                          & Cc~(P)                       & Fddd~(NP)              & 467.4 & 231.3 \\
	    V$_{2}$FeO$_{4}$               & C2/c~(NP)                    & P1~(P)                 & 527.2 & 230.6 \\
	    Li$_{2}$VO$_{3}$               & C2/m~(NP)                    & P1~(P)                 & 396.6 & 218.8 \\
	    KNO$_{2}$                      & Imm2~(P)                     & P2$_{1}$/c~(NP)        & 528.8 & 196.6 \\
	    MgFe$_{2}$(CoO$_{2}$)$_{4}$    & C2/m~(NP)                    & P1~(P)                 & 591.8 & 187.9 \\
	    LiCoSbO$_{4}$                  & Imma~(NP)                    & P4$_{3}$22~(P)         & 580.8 & 179.0 \\
	    Li$_{2}$Ni$_{3}$SnO$_{8}$      & P4$_{3}$2$_{1}$2~(P)         & P$\overline{1}$~(NP)   & 567.8 & 177.0 \\
	    Fe$_{3}$OF$_{5}$               & C2/m~(NP)                    & Pm~(P)                 & 362.8 & 175.5 \\
	    BeCl$_{2}$                     & I4$_{1}$/acd~(NP)            & I$\overline{4}$3m~(P)  & 406.3 & 171.1 \\
	    NaNO$_{3}$                     & R$\overline{3}$c~(NP)        & P1~(P)                 & 429.1 & 166.7 \\
	    LiMnBO$_{3}$                   & P$\overline{6}$~(P)          & P$\overline{1}$~(NP)   & 510.4 & 159.2 \\
	    Li$_{2}$GeF$_{6}$              & P4$_{2}$/mnm~(NP)            & P321~(P)               & 481.5 & 159.0 \\
	    CaCO$_{3}$                     & P2$_{1}$/c~(NP)              & P6$_{5}$22~(P)         & 370.4 & 157.7 \\
	    Bi(BO$_{2}$)$_{3}$             & P2$_{1}$/c~(NP)              & Pca2$_{1}$~(P)         & 411.3 & 154.6 \\
	    Li$_{3}$TiFe$_{3}$O$_{8}$      & P6$_{3}$mc~(P)               & R$\overline{3}$m~(NP)  & 313.0 & 152.9 \\
	    Li$_{5}$CoO$_{4}$              & Aea2~(P)                     & Pbca~(NP)              & 550.8 & 152.3 \\
	    Li$_{2}$MnSiO$_{4}$            & P$\overline{1}$~(NP)         & Pc~(P)                 & 595.9 & 152.0 \\
	& & & & \\
        \hline
        \hline
    \end{tabular}
    \caption{\textbf{Stable polar-nonpolar phase transformations predicted to exhibit a phase transition entropy 
	change higher than $150$~J~K$^{-1}$~kg$^{-1}$.} The low-$T$ polymorph corresponds to ``Phase 1'' and the 
	high-$T$ polymorph to ``Phase 2''.  ``P'' and ``NP'' stand for polar and nonpolar structures, respectively. 
	$T_{t}$ and $\Delta S_{t}$ stand for phase-transition temperature and entropy change, respectively.}
    \label{tab3}
\end{table*}

\begin{figure*}[t]
    \centering
    \includegraphics[width=1.0\linewidth]{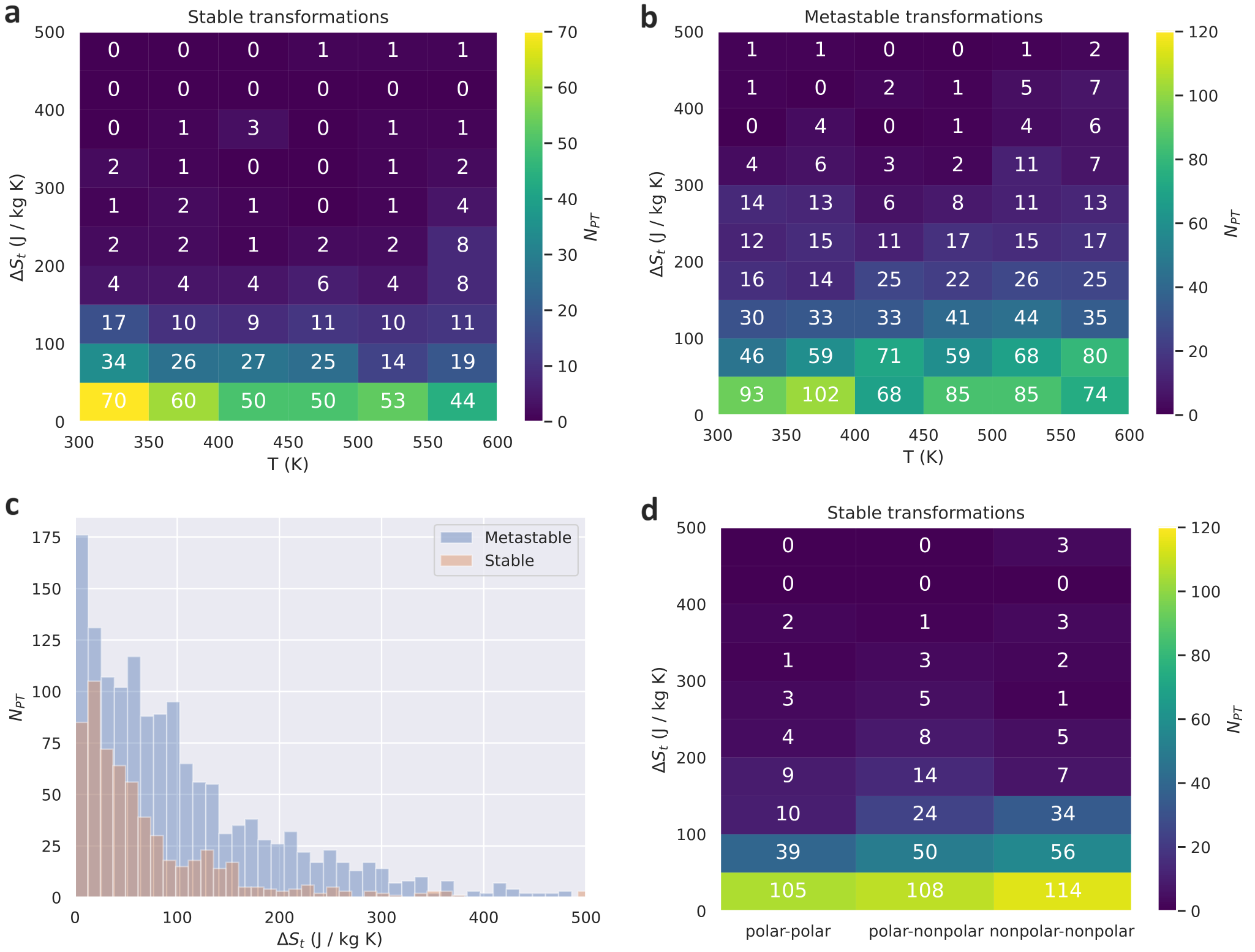}
	\caption{\textbf{Predicted phase-transition entropy change, $\Delta S_{t}$, expressed as a function of transition 
	temperature and crystal symmetry properties.}
        \textbf{a.}~Phase-transition entropy change for stable phase transitions as a function of temperature.
	\textbf{b.}~Phase-transition entropy change for metastable phase transitions as a function of temperature.
        \textbf{c.}~Number of phase transitions as a function of $\Delta S_{t}$.
	\textbf{d.}~Phase-transition entropy change for stable phase transitions as a function of symmetry.
	Numbers in white colour correspond to $N_{PT}$.
        }
    \label{fig6}
\end{figure*}

\subsection{Application~1: Solid-state cooling}
\label{subsec:app1}
Solid-state cooling methods represent energy-efficient and environmentally friendly alternatives to conventional 
refrigeration technologies, which typically rely on compression cycles involving greenhouse gases 
\cite{caloric1,caloric2,caloric3,caloric4}. Under moderate variations of magnetic, electric, or mechanical fields, 
promising caloric materials can undergo field-induced phase transitions that produce large adiabatic temperature 
changes ($|\Delta T| \sim 1$--$10$~K) and substantial isothermal entropy variations ($|\Delta S| \sim 
10$--$100$~J~K$^{-1}$~kg$^{-1}$). These caloric effects form the basis of solid-state refrigeration technologies, 
offering a pathway toward sustainable and high-performance cooling applications \cite{caloric5}. 

A schematic of a possible four-step refrigeration cycle based on the caloric effect is presented in Fig.\ref{fig5}a. 
The refrigeration cycle begins with the caloric material in a disordered (high-entropy) phase at room temperature, 
$T_{0}$. (This choice is arbitrary and serves illustrative purposes; other initialization conditions would also be 
valid.) An external field (e.g., hydrostatic pressure) is then adiabatically applied, inducing a transition to an 
ordered (low-entropy) phase, which raises the sample temperature to $T_{+} = T_{0} + |\Delta T|$. While the external 
field is maintained, the sample is brought into contact with a heat sink, enabling the release of heat ($+Q = 
|\Delta S|/T_{0}$) to the environment and returning the material to its initial temperature, $T_{0}$. The external 
field is then adiabatically removed, reverting the material to the disordered phase and lowering its temperature to 
$T_{-} = T_{0} - |\Delta T|$. In the final step, the sample is placed in thermal contact with the body to be cooled, 
absorbing heat and returning once more to $T_{0}$, thereby completing the cooling cycle.

From a practical perspective, achieving both large $|\Delta T|$ and $|\Delta S|$ is essential for the development of 
rapid and efficient solid-state cooling devices operating under repeated application and removal of external driving 
fields. The maximum amount of heat that can be extracted from a heat source per cycle is determined by $|\Delta S|$, 
while the rate at which this heat can be transferred to a heat sink is governed by $|\Delta T|$. To a first approximation, 
the adiabatic temperature change can be estimated as $\Delta T \approx -T_{0}/C \cdot \Delta S$, where $C$ is the heat 
capacity of the caloric material, and $\Delta S \approx \Delta S_{t}$ \cite{escorihuela24}. Therefore, the entropy 
change associated with a phase transition serves as a useful descriptor for assessing the caloric potential of a 
phase-change material.

Figure~\ref{fig6}a displays the total number of predicted stable phase transformations, $N_{PT}$, classified according 
to their transition temperature and entropy change. Regardless of temperature, higher values of $N_{PT}$ are generally 
correlated with smaller phase-transition entropy changes. For example, within the interval $300 \le T_{t} \le 350$~K, a 
total of $121$ stable phase transitions exhibit entropy changes equal to or below $150$~J~K$^{-1}$~kg$^{-1}$, whereas 
only $9$ transitions exceed this threshold. Conversely, the largest $\Delta S_{t}$ values are predominantly observed at 
higher temperatures. Notably, three phase transitions present exceptionally large entropy changes above 
$450$~J~K$^{-1}$~kg$^{-1}$; however, these transformations occur at temperatures exceeding $450$~K, which may limit their 
technological relevance for applications near room temperature.

Figure~\ref{fig6}b presents analogous information to that shown in Fig.\ref{fig6}a, but for metastable phase transitions. 
The general trends remain consistent with those observed for stable transformations: larger values of $N_{PT}$ are typically 
associated with smaller $\Delta S_{t}$, and larger entropy changes tend to occur at higher transition temperatures. For 
both stable and metastable cases, $N_{PT}$ decreases monotonically with increasing $\Delta S_{t}$, as illustrated in 
Fig.\ref{fig6}c. However, for any given $\Delta S_{t}$, the number of metastable phase transitions consistently exceeds 
the number of stable ones (Fig.\ref{fig6}c). As noted in Sec.\ref{subsec:general}, this trend can be attributed to the fact 
that considering multiple metastable polymorphs per each compound naturally increases the number of possible phase transitions.

Table~4 lists the stable phase transitions predicted to occur within the technologically relevant temperature range of 
$300 \le T_{t} \le 350$~K and that exhibit entropy changes equal to or greater than $100$~J~K$^{-1}$~kg$^{-1}$. These 
transformations are particularly promising for solid-state cooling applications, as they take place near room temperature 
and could potentially give rise to so-called ``giant'' and ``colossal'' caloric effects \cite{giant1,giant2,colossal1,colossal2}. 
Notably, the first five compounds in Table~4 display exceptionally large transition entropy changes: Li$_{3}$MnF$_{7}$ 
($342.3$~J~K$^{-1}$~kg$^{-1}$ at $347$~K), PNCl$_{2}$ ($338.3$~J~K$^{-1}$~kg$^{-1}$ at $349$~K), Fe$_{4}$OF$_{7}$ 
($252.1$~J~K$^{-1}$~kg$^{-1}$ at $300$~K), VOF$_{3}$ ($241.7$~J~K$^{-1}$~kg$^{-1}$ at $311$~K), and Mg$_{2}$TiO$_{4}$ 
($209.5$~J~K$^{-1}$~kg$^{-1}$ at $344$~K). To the best of our knowledge, the caloric properties of these compounds have 
not yet been experimentally explored. 

Figure~\ref{fig6}d shows the total number of predicted stable phase transformations, categorized by their entropy 
change and crystal symmetry type. When considering the entire temperature range of $300 \le T_{t} \le 600$~K, the 
number of phase transitions for a given $\Delta S_{t}$ is generally smallest for polar-polar transformations and 
largest for nonpolar-nonpolar phase transitions (in constrast to what we found near room temperature conditions, 
Sec.\ref{subsec:general}). Notably, only nonpolar-nonpolar transitions are predicted to exhibit entropy changes 
exceeding $400$~J~K$^{-1}$~kg$^{-1}$; however, the associated transition temperatures occur well above room 
temperature (Fig.\ref{fig6}a).

\begin{table*}[h]
    \centering
    \begin{tabular}{c c c c c c c c}
        \hline
        \hline
	    & & & & & & & \\
        \quad Compound \qquad & \quad Phase~1 \qquad &  \quad Phase~2 \qquad & \quad $T_{t}$~(K) \qquad &
	    \quad $\lambda_{1}$~(W~m$^{-1}$~K$^{-1}$) \qquad & \quad $ \lambda_{2}$~(W~m$^{-1}$~K$^{-1}$) \qquad &  
	    \quad $\Delta \lambda_{t} / \lambda_{1}$ \qquad & \quad \quad $\epsilon$ \qquad \quad \\
	    & & & & & & & \\
        \hline
	    & & & & & & & \\
	    ZrSeO                             & P2$_{1}$3~(P) 		& P4/nmm~(NP) 		& 311.7 & 8.1  & 14.0 & +0.73 & 1.73\\
	    VF$_{4}$                          & Pc~(P) 			& C2/c~(NP) 		& 486.5 & 2.5  & 3.6  & +0.44 & 1.44\\
	    V$_{2}$FeO$_{4}$                  & C2/c~(NP) 		& P1~(P) 		& 527.2 & 8.7  & 12.4 & +0.43 & 1.43\\
	    Li$_{4}$TiS$_{4}$                 & Pnma~(NP) 		& I$\overline{4}$2m~(P)	& 362.8 & 4.6  & 6.6  & +0.43 & 1.43\\
	    LiNd(MoO$_{4}$)$_{2}$             & I$\overline{4}$~(P) 	& P2/c~(NP)		& 566.0 & 2.8  & 3.8  & +0.36 & 1.36\\
	    Y$_{2}$Zr$_{2}$O$_{7}$            & P2$_{1}$~(P) 		& C2/m~(NP) 		& 592.0 & 6.2  & 8.3  & +0.34 & 1.34\\
	    Li$_{3}$Mn$_{2}$F$_{7}$           & Cc~(P) 			& P2$_{1}$/c~(NP) 	& 556.1 & 4.2  & 5.6  & +0.33 & 1.33\\
	    LiFeSbO$_{4}$                     & Imma~(NP) 		& P4$_{3}$22~(P) 	& 478.3 & 4.9  & 6.4  & +0.31 & 1.31\\
	    MgVFeP$_{2}$(O$_{4}$F)$_{2}$      & P$\overline{1}$~(NP) 	& P1~(P) 		& 478.1 & 4.7  & 5.9  & +0.26 & 1.26\\
	    LiMnBO$_{3}$                      & P$\overline{6}$~(P)	& P$\overline{1}$~(NP)	& 510.4 & 6.6  & 8.2  & +0.24 & 1.24\\
	    MgFe$_{4}$(PO$_{4}$)$_{4}$        & Pm~(P) 			& P$\overline{1}$~(NP) 	& 337.9 & 6.0  & 7.4  & +0.23 & 1.23\\
	    NaClO$_{4}$                       & Pnma~(NP) 		& I$\overline{4}$2m~(P)	& 538.3 & 3.2  & 3.9  & +0.22 & 1.22\\
	    Li$_{3}$V(Si$_{2}$O$_{5}$)$_{3}$  & P1~(P) 			& Cmce~(NP) 		& 358.3 & 11.7 & 9.3  & -0.20 & 1.26\\
	    Li$_{3}$CoPCO$_{7}$               & P2$_{1}$~(P) 		& P2$_{1}$/m~(NP) 	& 322.5 & 5.3  & 4.2  & -0.20 & 1.26\\
	    Na$_{3}$SrPCO$_{7}$               & P2$_{1}$/m~(NP) 	& P2$_{1}$~(P) 		& 396.1 & 3.0  & 2.4  & -0.20 & 1.25\\
	    Ba$_{2}$MgCrMoO$_{6}$             & Immm~(NP)		& Cm~(P)		& 313.5 & 3.5  & 2.8  & -0.21 & 1.25\\
	    Li$_{2}$MnSiO$_{4}$               & P$\overline{1}$~(NP) 	& Pc~(P)  		& 595.9 & 9.4  & 7.4  & -0.22 & 1.27\\
	    Y$_{2}$Zr$_{2}$O$_{7}$            & Fd$\overline{3}$m~(NP) 	& P2$_{1}$~(P) 		& 323.6 & 8.0  & 6.2  & -0.23 & 1.29\\
	    NaNO$_{3}$                        & R$\overline{3}$c~(NP) 	& P1~(P) 		& 429.1 & 5.3  & 3.6  & -0.31 & 1.47\\
	    Bi$_{2}$W$_{2}$O$_{9}$            & Pna2$_{1}$~(P) 		& Pbcn~(NP) 		& 346.5 & 4.0  & 2.3  & -0.41 & 1.74\\
	    CoO$_{2}$                         & P1~(P) 			& C2/m~(NP) 		& 592.3 & 11.7 & 6.2  & -0.47 & 1.89\\
	    & & & & & & & \\
        \hline
        \hline
    \end{tabular}
    \caption{\textbf{Stable polar-nonpolar phase transformations predicted to exhibit a relative heat conductivity change 
	higher than $0.2$ (in absolute value).} The low-$T$ polymorph corresponds to ``Phase 1'' and the high-$T$ polymorph 
	to ``Phase 2''.  ``P'' and ``NP'' stand for polar and nonpolar structures, respectively. $T_{t}$, $\Delta S_{t}$
	and $\Delta \lambda_{t}$ stand for phase-transition temperature, entropy change and heat conductivity change, 
	respectively. $\epsilon$ is defined as the ratio of the high LTC (``on'') state by the low LTC (``off'') state.}
    \label{tab4}
\end{table*}

Finally, Table~5 lists all stable polar-nonpolar phase transitions predicted to exhibit entropy changes equal to 
or greater than $150$~J~K$^{-1}$~kg$^{-1}$, regardless of the transition temperature. This classification is 
particularly useful for identifying materials with the potential to exhibit large electrocaloric effects, which 
are induced by variations in electric fields \cite{cazorla18,giant1,colossal2}. At the top of the list is the 
polymeric compound PNF$_{2}$, which displays a highly promising $\Delta S_{t}$ of $360.7$~J~K$^{-1}$~kg$^{-1}$ and 
a moderately high transition temperature of $411$~K. It is followed by the neutral salt KNO$_{3}$, which exhibits 
an impressive entropy change of $346.4$~J~K$^{-1}$~kg$^{-1}$, although its high transition temperature of $539$~K 
may limit its suitability for conventional refrigeration applications. Among the entries in Table~5, we highlight 
several materials with transition temperatures near ambient conditions: PNCl$_{2}$ ($338.3$~J~K$^{-1}$~kg$^{-1}$), 
Cd(C$_{4}$N$_{3}$)$_{2}$ ($259.0$~J~K$^{-1}$~kg$^{-1}$), Li$_{2}$VO$_{3}$ ($218.8$~J~K$^{-1}$~kg$^{-1}$), 
Fe$_{3}$OF$_{5}$ ($175.5$~J~K$^{-1}$~kg$^{-1}$), CaCO$_{3}$ ($157.7$~J~K$^{-1}$~kg$^{-1}$), and Li$_{3}$TiFe$_{3}$O$_{8}$ 
($152.9$~J~K$^{-1}$~kg$^{-1}$).

Overall, in this section we have provided valuable insights into materials that, to the best of our knowledge, have 
not yet been explored from the perspective of solid-state refrigeration. Owing to their large predicted entropy changes 
and favorable transition temperatures, these compounds emerge as promising candidates for advancing the development of 
next-generation solid-state cooling technologies.

\begin{figure*}[t]
    \centering
    \includegraphics[width=1.0\linewidth]{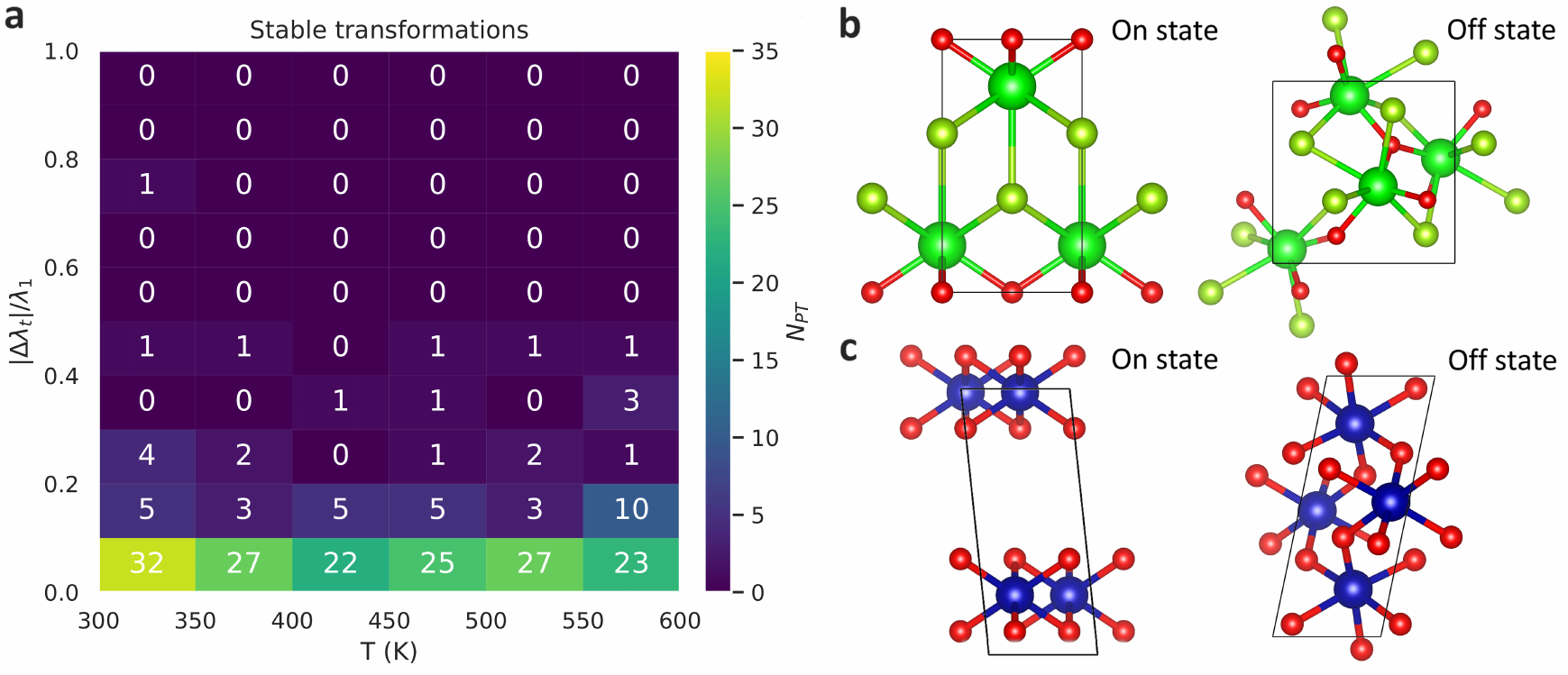}
	\caption{\textbf{Results obtained for the proposed thermal switch application.}
	\textbf{a.}~Number of polar-nonpolar phase transitions, $N_{PT}$, expressed as a function of relative 
	thermal conductivity change (in absolute value) and transition temperature.
	\textbf{b.}~Ball-stick representation of the ``on'' and ``off'' thermal states of ZrSeO which, according
	to our ML-guided high-throughput screening, exhibits a relative LTC change of approximately 73\% (Table~6). 
	Zr, Se and O ions are represented with green, yellow-green and red spheres, respectively.
	\textbf{c.}~Ball-stick representation of the ``on'' and ``off'' thermal states of CoO$_{2}$ which, according 
	to our ML-guided high-throughput screening, exhibits a relative LTC change of approximately 47\% (Table~6). 
	Co and O ions are represented with blue and red spheres, respectively.}
    \label{fig7}
\end{figure*}

\subsection{Application~2: Thermal switching}
\label{subsec:app2}
A thermal switch is a device or material system that exhibits a controllable and reversible change in thermal 
conductivity in response to external stimuli such as temperature, electric fields, magnetic fields, light, or 
mechanical stress \cite{ts0,ts1,ts2,ts3,ts4,ts5}. The fundamental objective of a thermal switch is to regulate 
heat flow in a precise and dynamic manner, analogous to how an electrical switch controls current (Fig.\ref{fig5}b). 
Technically, a thermal switch can be characterized by its thermal conductance ratio, defined as the ratio of 
thermal conductivity in the ``on'' (high-conductance) state to that in the ``off'' (low-conductance) state, 
$\epsilon = \lambda_{\rm on} / \lambda_{\rm off}$. Thermal switches are important in numerous advanced technological 
fields, including thermal management in electronics, energy conversion systems (e.g., thermoelectric generators) 
and neuromorphic and phononic devices \cite{ts6}. 

Thermal switches achieve modulation of heat flow by altering the way phonons (quantized lattice vibrations) or 
electrons (in conductive materials) transport energy. Key mechanisms on which thermal switches may rely include 
anharmonic phonon scattering (e.g., activation of low-frequency rattling modes) \cite{ts0}, interfacial thermal 
resistance modulation (e.g., phase-change layers that alternately enhance or block interfacial heat transfer) 
\cite{ts1}, external field control of lattice properties leading to changes in thermal transport \cite{ts2}, and 
phase transitions (e.g., order-disorder, metal-insulator and structural transformations) \cite{ts3,ts4,ts5}.

In this study, we select phase transitions as the underlying mechanism for proposing auspicious thermal switches. 
Specifically, by employing a surrogate ML model capable of predicting a material's lattice thermal conductivity 
(LTC) based solely on its composition and atomic structure \cite{ltc-ml1,ltc-ml2,ltc-ml3,ltc-ml4} (Methods), we 
systematically investigated the relative LTC changes across all the stable polar-nonpolar phase transitions predicted 
in this work, totaling $214$ transformations (Fig.\ref{fig4}a). This class of transitions was selected because 
they offer the possibility to be externally controlled via an applied electric field.

Figure~\ref{fig7}a summarizes the outcomes of our ML-guided high-throughput screening of LTC, showing the number 
of phase transitions, $N_{PT}$, as a function of transition temperature and relative LTC change (in absolute value), 
$|\Delta \lambda_{t}|/\lambda_{1}$ (with $\lambda_{1}$ and $\lambda_{2}$ denoting the LTC of the low- and high-$T$ 
phases, respectively). We find that approximately 75\% of the predicted stable polar-nonpolar phase transitions 
exhibit relative LTC changes equal to or smaller than $0.1$. Only $31$ transformations display variations in the 
$0.1$--$0.2$ range, and just $21$ exceed $0.2$. For any given $|\Delta \lambda_{t}|/\lambda_{1}$, the distribution 
of $N_{PT}$ across the $300 \le T_{t} \le 600$~K temperature range remains approximately uniform.

Table~6 lists the $21$ stable polar-nonpolar phase transitions predicted to exhibit relative LTC changes equal to or 
greater than $0.20$ in absolute value. At the top of the list is the Janus compound ZrSeO \cite{zrseo} (Fig.\ref{fig7}b), 
which displays a remarkable $\Delta \lambda_{t}/\lambda_{1}$ value of $+0.73$ and a transition temperature of $312$~K, 
ideal for near-room-temperature applications. The corresponding LTC ratio, $\epsilon$, is also notably large, with a 
value of $1.73$. At the bottom of the list is the binary oxide CoO$_{2}$ (Fig.\ref{fig7}c), a typical cathode material 
in lithium-based electrochemical batteries \cite{zhou23}, which features a significant $\Delta \lambda_{t}/\lambda_{1}$ 
value of $-0.47$ and the highest $\epsilon$ value in the table, $1.89$. However, its high transition temperature in 
principle suggests limited applicability in conventional thermal management technologies.

Other notable cases in Table~6 correspond to phase transitions occurring near room temperature, which are particularly 
relevant for practical thermal switching applications. These include Li$_{4}$TiS$_{4}$ with $\Delta \lambda_{t}/\lambda_{1} 
= +0.43$ and $\epsilon = 1.43$, MgFe$_{4}$(PO$_{4}$)$_{4}$ with $+0.23$ and $1.23$, Li$_{3}$V(Si$_{2}$O$_{5}$)$_{3}$ and 
Li$_{3}$CoPCO$_{7}$ with $-0.20$ and $\epsilon = 1.26$ both, Ba$_{2}$MgCrMoO$_{6}$ with $-0.21$ and $1.25$, 
Y$_{2}$Zr$_{2}$O$_{7}$ with $-0.23$ and $1.29$, and Bi$_{2}$W$_{2}$O$_{9}$ with $-0.41$ and an impressive $\epsilon = 1.74$. 
These materials, along with ZrSeO and CoO$_{2}$, stand out as promising candidates for the development of efficient thermal 
switches, which, to the best of our knowledge, have not yet been experimentally investigated.

\section{Discussion}
\label{sec:discussion}
Given the generally complex compositions and low crystal symmetries of the compounds listed in Tables~1--6, validating 
our temperature-induced phase transition predictions directly with first-principles QHA calculations turns out to be 
very challenging. Nevertheless, for a reduced subset of these materials, we are able to compare our ML-aided predictions 
with available experimental data. This is the case for the ternary oxides KNO$_3$, KNO$_2$, and CaCO$_3$ (Table~5), 
which have been previously analyzed and reported in the literature. 

For bulk KNO$_3$, a ferroelectric phase transition has been experimentally observed within the temperature range $350 
\le T_{t} \le 400$~K, depending on the cooling rate and sample preparation conditions \cite{kno1,kno2}. Our high-throughput 
screening predicts a transition temperature of $539$~K for this compound, which is reasonably close to the experimental range. 
Likewise, for KNO$_2$, a polar phase transition has been reported at $313$~K \cite{kno}, which also compares reasonably 
with our predicted value of $529$~K. In the case of slightly compressed CaCO$_3$, a phase transition has been observed at 
$336$~K \cite{caco}, which is in very good agreement with our estimate of $370$~K. This overall consistency between predicted 
and experimental transition temperatures supports the reliability of our computational approach and strengthens confidence in 
the broader set of predictions presented in this work.

For a reduced subset of $10$ predicted stable phase transitions, which neither involve chemically complex nor low-symmetry 
structures, we have been able to perform theoretical validation at the first-principles level using DFT (Methods). The results 
of these DFT validation tests are presented and discussed in the Supplementary Information and Supplementary Discussion. Overall, 
for materials that do not exhibit imaginary phonon frequencies at zero temperature, we find good agreement among the ML-predicted 
and DFT-calculated free-energy differences between the two polymorphs involved in the phase transition, with discrepancies well 
within the expected numerical uncertainties.

Additionally, we have computed the thermal conductance ratio using first-principles DFT methods, $\epsilon^{\rm DFT}$, 
for two representative compounds exhibiting polar-nonpolar phase transformations, Li$_{4}$TiS$_{4}$ and NaNO$_{3}$, both 
of which are vibrationally stable at zero temperature (Methods). For Li$_{4}$TiS$_{4}$, we obtained $\epsilon^{\rm DFT} 
= 2.23$, and for NaNO$_{3}$, $\epsilon^{\rm DFT} = 1.40$. These results compare notably well with the ML-predicted values 
of $1.43$ and $1.47$, respectively (Table~6). This level of consistency supports the overall reliability of our ML-guided 
phase-transition and lattice thermal conductivity predictions.

Despite the encouraging results presented in this work, our high-throughput screening strategy is subject to a number 
of inherent limitations. First, thermal expansion effects have been systematically neglected throughout our analysis. 
Since our transition temperature estimates rely on the quasi-harmonic approximation applied to zero-temperature relaxed 
structures, discrepancies with respect to available experimental data may arise, especially in cases where thermal 
expansion plays a significant role in the relative phase stability. Incorporating anharmonic effects or performing 
fully temperature-dependent structural relaxations could improve the accuracy of transition temperature predictions, 
albeit at a significantly higher computational cost \cite{lopez23,lopez24,benitez25}. The development of subsidiary ML 
models and/or the use of ML interatomic potentials could enormously facilitate such a materials modeling endeavour
\cite{walsh23}.

Second, our approach is fundamentally limited to solid-solid phase transitions between structurally well-defined atomic 
crystals. If a material undergoes a different kind of temperature-induced transformation in practice (e.g., melting, 
sublimation, or a transition to a dynamically disordered superionic state) our methodology, which is based on quasi-harmonic 
phonon calculations, will not be able to identify it. Consequently, some compounds predicted to exhibit stable solid-solid 
transitions may instead experience a different transformation pathway upon heating, which should be taken into account 
when considering experimental validation. 

Finally, the limited scope of the employed materials dataset \cite{mp} also constrains our findings. Promising compounds 
not included in this repository could not be captured by our screening, highlighting the importance of continually expanding 
open-access materials databases to support future discoveries. In this context, generative crystal structure models based 
on machine learning and data-driven approaches offer a powerful route to explore chemically and structurally diverse materials 
spaces beyond what has been experimentally and/or theoretically investigated. Integrating such generative tools with 
high-throughput workflows could significantly enhance the discovery potential of future computational screenings \cite{matergen}.

\section{Conclusions}
\label{sec:conclusions}
We have developed and implemented a ML-guided high-throughput framework for predicting temperature-induced solid-solid 
phase transitions in inorganic materials. By integrating DFT calculations with a graph-based neural network trained to 
estimate vibrational free energies, we have successfully screened a large dataset of materials from the Materials Project
and identified over $2,000$ potential polymorphic transformations within the temperature range of $300$--$600$~K. Our 
method enables a significant computational speed-up compared to traditional first-principles approaches, while maintaining 
predictive accuracy in line with available experimental data and fully \textit{ab initio} calculations.

The systematic analysis of these phase transitions revealed a rich variety of functional phenomena with potential 
for technological applications. We identified hundreds of polar-nonpolar and polar-polar transitions near room 
temperature, many of which exhibit large entropy changes ($> 300$~J~K$^{-1}$~kg$^{-1}$) and are therefore promising 
candidates for caloric and solid-state refrigeration applications. Additionally, we reported $21$ materials with sharp 
relative changes in lattice thermal conductivity ($20$--$70$\%) across a phase transition, offering a new class of 
thermally switchable materials. Notably, several of the highlighted compounds --such as ZrSeO, Bi$_{4}$W$_{2}$O$_{9}$, 
and Li$_{4}$TiS$_{4}$-- have not previously been investigated for these applications, and thus represent new 
opportunities for experimental exploration. 

Our ML-guided high-throughput screening approach for investigating solid-solid phase transitions in inorganic materials 
represents a transformative methodological advancement with profound implications for thermal management technologies. 
Future research will strategically expand our methodology to organic materials and energy applications. The proposed 
methodology establishes a robust, scalable, and extensible computational framework for phase-transition-driven materials 
exploration, particularly when integrated with cutting-edge generative AI models and expansive materials databases.

\section*{Methods}
\label{sec:methods}
\textbf{First-principles DFT calculations.}~\textit{Ab initio} calculations based on density functional theory (DFT) 
\cite{cazorla17} were performed to analyse the structural and vibrational properties of selected materials. We performed 
these calculations with the VASP code \cite{vasp} using the semi-local PBEsol \cite{pbesol} approximation to the 
exchange-correlation energy functional. The projector augmented-wave (PAW) method was used to effectively model the ionic 
cores \cite{bloch94}. Wave functions were represented in a plane-wave basis typically truncated at $650$~eV. By using 
these parameters and dense {\bf k}-point grids for reciprocal-space integration the resulting zero-temperature energies 
were converged to within $1$~meV per formula unit. In the geometry relaxations, a force tolerance of $0.005$~eV$\cdot$\AA$^{-1}$ 
was imposed in all the atoms. The second-order interatomic force constant (IFC) matrix of selected materials were calculated 
with the finite-differences method as is implemented in the \verb!PhonoPy! software \cite{phonopy}. Large supercells ($4 
\times 4 \times 4$) and dense {\bf k}-point grids were employed for the phonon and vibrational free-energy calculations of 
targeted structures. 
\\

\textbf{Vibrational free-energy ML model.}~To train our GCNN for predicition of vibrational free energies, we used 
a supervised learning approach in which a task-specific loss function was defined (i.e., mean squared error for 
regression) and optimized using backpropagation and gradient descent. Specifically, the GCNN was implemented with 
three convolutions of node features, batch normalization and average pooling, and two further layers of convolution, 
with ReLU as activation functions and Adam optimizer. We included a batch normalization layer in between the graph 
convolutional layers for improving model stability.

To evaluate the role of hyperparameters in our GCNN, we performed random hyperparameter search \cite{Bergstra2012} 
on small training sets of $80$ epochs with the following grid values: graph convolutional layers with $[64, 128, 256, 
512]$ number of neurons, linear convolutional layers with $[16, 32, 64, 128, 256]$ number of neurons, dropouts of 
$[0.1, 0.2, 0.3, 0.4, 0.5]$ and learning rates of $[0.00005, 0.0001, 0.0005, 0.001]$. Upon selection of the best
hyperparameters, longer trainings were performed to improve subsequent screenings and postprocessings.

Finally, an optimal GCNN vibrational free-energy model was obtained consisting of two graph convolutional layers 
of $512$ neurons each, two linear convolutional layers with $64$ and $16$ neurons each, and a linear output. The 
learning rate was set to $0.001$, dropout to $0.1$ and batch size to $64$.
\\

\textbf{First-principles LTC calculations.}~The DFT calculations were performed with the VASP code \cite{vasp} using 
the semi-local PBEsol \cite{pbesol} approximation to the exchange-correlation energy functional and PAW method \cite{bloch94}. 
The kinetic energy cutoff of the plane wave basis set is set to be $520$~eV. Convergence thresholds of the total 
energy and Hellmann-Feynman force were set to $10^{-8}$~eV and $10^{-4}$~eV/\AA, respectively. The second-order 
IFC matrix of selected materials were calculated with the finite-differences method as is implemented in the \verb!PhonoPy! 
software \cite{phonopy}. Large supercells containing between $100$ and $300$ atoms were employed in these calculations. 
For the third-order IFC matrix calculations, the interatomic interactions were truncated to the third nearest neighbor. 
Finally, using the 2nd and 3rd IFCs as inputs, the phonon Boltzmann transport equation was iteratively solved using the 
\verb!ShengBTE! package \cite{shengbte}. The NGRIDS parameter in the \verb!ShengBTE! calculations was large enough to 
ensure the total number of phonon scattering channels to be of the order of $10^{8}$.
\\

\textbf{LTC subsidiary ML model.}~Atomistic Line Graph Neural Network (ALIGNN) \cite{alignn} was introduced as a direct 
ML model, whereby the inputs are automatically calculated based on the structural and chemical composition of crystals, 
where the former includes both bond distances and angles. In total, $\approx 4,700$ LTC data calculated by full DFT 
calculations were used as end property for ALIGNN model training, which were accumulated in our recent works 
\cite{ltc-ml1,ltc-ml2,ltc-ml3,ltc-ml4}. The logscale LTC values are used in the ALIGNN model training, as our separate 
tests have proved significant performance improvement as compared to the models that were trained on the raw LTC values. 
Also, for anisotropic structures whose LTC values were different in the three crystallographic directions, a single 
LTC value averaged over all three directions was used for training. The ALIGNN model was trained on 80\% of the dataset 
and the rest 20\% was used for testing. The model was trained for $2,000$ epochs with a batch size of $100$.
\\

\section*{Data availability}
The data that support the findings of this study are freely available at \cite{github}. 
\\

\section*{Acknowledgements}
C.L. acknowledges support from the Spanish Ministry of Science, Innovation and Universities under a FPU grant. 
C.C. acknowledges support by MICIN/AEI/10.13039/501100011033 and ERDF/EU under the grants TED2021-130265B-C22,
TED2021-130265B-C21, PID2023-146623NB-I00, PID2023-147469NB-C21 and RYC2018-024947-I and by the Generalitat de 
Catalunya under the grants 2021SGR-00343, 2021SGR-01519 and 2021SGR-01411. Computational support was provided 
by the Red Española de Supercomputación under the grants FI-2024-1-0005, FI-2024-2-0003, FI-2024-3-0004,
FI-2024-1-0025, FI-2024-2-0006, and FI-2025-1-0015. This work is part of the Maria de Maeztu Units of Excellence 
Programme CEX2023-001300-M funded by MCIN/AEI (10.13039/501100011033). E.S. acknowledges the European Union H2020 
Framework Program SENSATE project: Low dimensional semiconductors for optically tuneable solar harvesters (grant 
agreement Number 866018), Renew-PV European COST action (CA21148) and the Spanish Ministry of Science and Innovation 
ACT-FAST (PCI2023-145971-2). E.S. and J.-Ll.T. are grateful to the ICREA Academia program. M.H. acknowledges 
support from NSF (Award no. 2030128).
\\

\end{document}